\documentclass[10pt,a4letter]{article}
\usepackage[english]{babel}
\usepackage{comment}

\usepackage{amsmath}
\usepackage{amssymb}
\usepackage{dsfont}
\usepackage{graphicx,epsfig}
\usepackage{slashed}
\usepackage[mathscr]{eucal}
\usepackage{placeins}

\usepackage{color}
\numberwithin{equation}{section}

\usepackage[numbers,sort&compress]{natbib}
\usepackage{multirow}
\usepackage{tabularray}
\usepackage{tabularx}
\UseTblrLibrary{booktabs}
\usepackage{float}

\usepackage{slashed}
\usepackage{braket}

\usepackage{scalefnt,ulem,pstricks}

\usepackage{xspace}
\usepackage{setspace}
\usepackage{xstring}
\usepackage{mathrsfs}
\usepackage{amsbsy}
\usepackage{makecell}

\usepackage{lipsum}
\usepackage[colorlinks=true
,urlcolor=blue
,anchorcolor=blue
,citecolor=blue
,filecolor=blue
,linkcolor=blue
,menucolor=blue
,pagecolor=blue
,linktocpage=true
,pdfproducer=medialab
,pdfa=true
]{hyperref}
\usepackage[capitalise]{cleveref}
\usepackage{shuffle}
\usepackage{subfig}

\newcommand\asbare{\alpha_s^0}
\newcommand\as{\alpha_s}
\newcommand\MSbar{\overline{\mathrm{MS}}}

\newcommand\M{{\cal M}}
\newcommand\A{{\cal A}}
\newcommand\F{{\cal F}}
\newcommand\R{{\cal R}}
\newcommand\Hard{{\cal H}}

\newcommand\OpenLoops{{\sc OpenLoops}\xspace}

\newcommand{\eps}{\epsilon}

\newcommand{\ii}{\mathrm{i}}
\newcommand{\dd}{{\mathrm{d}}}

\def\ttH{\ensuremath{t \bar t H}\xspace}
\def\ttW{\ensuremath{t \bar t W}\xspace}

\begin{document}
\begin{titlepage}
\begin{flushright}
ZU-TH 27/26 \\
TUM-HEP-1610/26 \\
MPP-2026-132\\
CERN-TH-2026-183
\end{flushright}

\vspace*{0.5cm}

%%%%%%%%%%%%%%%%%%%%%%%%%%%
\begin{center}
  {\Large \bf Two-loop QCD amplitudes for $\ttW$ production at the LHC \\ \vspace{0.2cm} in the leading-colour approximation}
\end{center}

\par \vspace{2mm}
\begin{center}
{\bf Matteo~Becchetti${}^{(a)}$}, {\bf Dhimiter~Canko${}^{(a)}$}, {\bf Xiang~Chen${}^{(b)}$}, {\bf Vsevolod~Chestnov${}^{(c)}$}, \\[0.2cm]
{\bf Maximilian~Delto${}^{(d)}$}, {\bf Sara~Ditsch${}^{(e,f)}$}, {\bf Tiziano~Peraro${}^{(a)}$}, {\bf Mattia~Pozzoli${}^{(a)}$}, \\[0.2cm]
{\bf Chiara~Savoini${}^{(f)}$} and {\bf Simone~Zoia${}^{(b)}$}
\vspace{5mm}\\
${}^{(a)}$ Dipartimento di Fisica e Astronomia, Universit\`{a} di Bologna and\\ INFN, Sezione di Bologna, via Irnerio 46, 40126 Bologna, Italy\\[0.25cm]
${}^{(b)}$ Physik-Institut, Universit\"at Z\"urich, Winterthurerstrasse 190, 8057 Z\"urich, Switzerland\\[0.25cm]
${}^{(c)}$ Mathematical Institute, University of Oxford, OX2 6GG, United Kingdom\\[0.25cm]
${}^{(d)}$ Theoretical Physics Department, CERN, 1211 Geneva 23, Switzerland\\[0.25cm]
${}^{(e)}$ Max-Planck-Institut f\"ur Physik, Werner-Heisenberg-Institut, Boltzmannstraße 8, 85748 Garching, Germany \\[0.25cm]
${}^{(f)}$ Technical University of Munich, TUM School of Natural Sciences, Physics Department, James-Franck-Straße 1, 85748 Garching, Germany\\[0.25cm]
\end{center}

%%%%% abstract %%%%%
\par \vspace{2mm}
\begin{center} {\large \bf Abstract}

\end{center}
\begin{quote}
\pretolerance 10000

We present a numerical computation of the two-loop QCD scattering amplitudes for the production of a top--antitop quark pair in association with a $W$ boson (\ttW) at the LHC in the generalised leading-colour approximation, retaining the exact dependence on the top-quark and $W$-boson masses. 
Rather than pursuing a fully analytic calculation, we employ a hybrid framework that combines numerical evaluation with strong algebraic and analytic control, allowing ultraviolet and infrared singularities as well as large intermediate cancellations to be treated exactly. 
This is achieved by expressing the finite remainder in terms of a set of special functions with rational coefficients. 
The special functions are evaluated numerically by solving differential equations through power-series expansions, while the values of the rational coefficients are reconstructed, point by point, from finite-field evaluations.
The calculation is performed in the 't Hooft-Veltman scheme and validated against an independent implementation in conventional dimensional regularisation employing a substantially different computational strategy. 
We finally provide the colour- and polarisation-summed hard functions evaluated on the phase-space grid used in a previous computation of the next-to-next-to-leading-order QCD corrections to the \ttW cross section.

\end{quote}

\vspace*{\fill}
August 2026
\end{titlepage}

\tableofcontents

%%%%%%%%%%%%%%%%%%%%%%%%%%
\section{Introduction}
\label{sec:introduction}

The precision reached by measurements at the Large Hadron Collider (LHC), together with the expected improvements during the High-Luminosity LHC phase, calls for theoretical predictions of unprecedented accuracy. 
In particular, next-to-next-to-leading-order (NNLO) QCD calculations have become essential for a range of $2\to3$ and higher multiplicity scattering processes. Their completion, however, is hindered by the need for multi-scale two-loop scattering amplitudes, whose complexity grows rapidly with the number of external particles and independent mass scales.
Among the processes requiring such calculations, those involving top quarks play a particularly important role. Owing to its large mass and correspondingly large Yukawa coupling, the top quark provides a unique probe of the electroweak (EW) symmetry-breaking mechanism. Moreover, top-quark production processes constitute an important background to many precision measurements within the Standard Model (SM) and to searches for physics beyond it.

In this work, we focus on the on-shell production of a top--antitop quark pair in association with a $W$ boson and compute the corresponding two-loop amplitudes in the generalised leading-colour approximation.
The \ttW process plays an important role in the LHC physics programme. 
It gives rise to distinctive multilepton final states, including same-sign dilepton signatures, which are relevant for new physics searches~\cite{ATLAS:2018alq,ATLAS:2019fag,CMS:2020cpy}. 
It also constitutes an irreducible background to several LHC analyses, most notably top--antitop production in association with a Higgs boson (\ttH) and four-top production. 
Moreover, measurements performed by the ATLAS and CMS collaborations have generally reported \ttW production rates exceeding the corresponding SM predictions, both in dedicated \ttW measurements~\cite{CMS:2022tkv,ATLAS:2024moy,CMS:2025iwa} and in indirect determinations~\cite{ATLAS:2019nvo,CMS:2020mpn,CMS:2023ftu,ATLAS:2023ajo}.
These considerations make increasingly precise theoretical predictions for \ttW production particularly important.

From the theoretical perspective, next-to-leading-order (NLO) QCD corrections to on-shell $\ttW$ production were computed in Refs.~\cite{Badger:2010mg,Campbell:2012dh,Maltoni:2015ena}, while the NLO EW corrections were evaluated in Refs.~\cite{Frixione:2015zaa,Frederix:2017wme}. 
The complete off-shell process was first computed at NLO QCD in Refs.~\cite{Bevilacqua:2020pzy,Denner:2020hgg,Bevilacqua:2020srb}, and subsequently including the full tower of EW corrections in Ref.~\cite{Denner:2021hqi}. 
NLO QCD+EW predictions have also been supplemented with multi-jet merging~\cite{Frederix:2012ps,Frederix:2021agh}, thus providing the theory reference for experimental measurements over the past several years.
More recently, a multi-scale improved NLO QCD calculation has been presented for the full off-shell process, including merged predictions with up to two additional jets~\cite{Dimitrakopoulos:2026jwi}.
In Ref.~\cite{Buonocore:2023ljm}, the inclusive on-shell $\ttW$ cross section was computed for the first time at next-to-next-to-leading order (NNLO) in QCD using two complementary dynamical approximations for the double-virtual contribution: the soft-$W$ and the high-energy approximations, respectively, where the latter neglects power corrections in the top-quark mass.
An exact calculation which retains the full kinematic dependence and includes genuine two-loop amplitudes is, however, required to validate these approximations and to assess their applicability beyond inclusive observables, ultimately enabling reliable NNLO predictions for differential distributions.

The presence of an external massive $W$ boson, together with massive quarks appearing both as external particles and in internal loops, makes the evaluation of the two-loop amplitudes particularly challenging.
The complexity of the calculation arises from two main sources.
First, the kinematics involves a large number of independent variables, leading to substantial algebraic complexity throughout the calculation and imposing stringent efficiency requirements on the numerical evaluation of the two-loop amplitude within a Monte Carlo integration framework.
Second, the analytic structure of the relevant loop integrals, in the presence of massive internal lines, typically involves higher-genus geometries, such as elliptic curves, which cannot be treated with the well-established methods widely used for massless $2 \to 3$ processes.
Together, these features make it challenging, with current techniques, to perform a fully analytic computation of such amplitudes and to efficiently evaluate them numerically.
For these reasons, despite their phenomenological importance, results for two-loop $2\to 3$ amplitudes with exact dependence on internal masses remain scarce.
Indeed, besides partial results for $\ttH$ production~\cite{FebresCordero:2023pww,Agarwal:2024jyq}, only the leading-colour two-loop QCD amplitudes for top-antitop production in association with a jet ($t\bar{t}j$) are currently available~\cite{Badger:2025ljy,Badger:2024dxo}.
The two-loop $\ttW$ amplitude, however, presents a significantly higher degree of both algebraic and analytic complexity than $t\bar{t}j$.
This led us to explore a different path.
In this paper, we adopt a hybrid strategy that allows us to evaluate the two-loop leading-colour amplitude for $\ttW$ production numerically, without deriving a fully analytic result, while still retaining strong analytic and algebraic control throughout the calculation.

The analytic complexity originates from the relevant loop integrals.
A complete set of master integrals (MIs), together with the differential equations (DEs) they obey, was obtained in Ref.~\cite{Becchetti:2025qlu}.
Building upon these results, we construct a set of special functions to represent the MIs following the method of Ref.~\cite{Badger:2024dxo}, which extends the framework widely applied to two-loop $2\to 3$ integrals with massless internal lines~\cite{Gehrmann:2018yef,Chicherin:2020oor,Chicherin:2021dyp,Abreu:2023rco}.
This construction does not require the DEs for the MIs to be cast into canonical form~\cite{Henn:2013pwa}, a step that typically underpins analytical methods but becomes particularly challenging in the presence of higher-genus geometries combined with the intricate $2\to 3$ kinematics.
The trade-off is that we may miss some relations among the functions associated with the most complicated analytic features: elliptic curves and nested square roots.
Nevertheless, this representation enables the exact cancellation of the ultraviolet (UV) and infrared (IR) poles of the two-loop amplitude, thereby allowing for the direct and simplified extraction of the finite remainder. 
We express this finite remainder in terms of linearly independent rational coefficients and special functions, and evaluate the latter numerically by solving the corresponding systems of DEs with \textsc{AMFlow}'s solver~\cite{Liu:2022chg}.

The key feature of our approach is that, rather than computing the rational coefficients analytically, we reconstruct their values exactly, point by point, from finite-field evaluations~\cite{vonManteuffel:2014ixa,Peraro:2016wsq} on a grid of rationalised phase-space points. 
The resulting grid was then interpolated to obtain the double-virtual contribution to the NNLO QCD $\ttW$ cross section in Ref.~\cite{Becchetti:2026awn}.
This strategy allows us to avoid the precision loss inherent in floating-point numerical calculations and to benefit from the important cancellations and simplifications that follow from the representation of the MIs in terms of special functions, without undertaking an expensive fully analytic computation.
While the point-wise reconstruction of rational coefficients had already been proposed (see, e.g., Ref.~\cite{Peraro:2019okx}, which we follow here), this is the first time that this method has been employed to cover the entire phase space and obtain an interpolation of the two-loop amplitude that is suitable for a phenomenological study.

We cross-check our results against an independent computational framework in a different regularisation scheme.
This alternative strategy directly targets the interference between the two-loop finite remainder and the Born amplitude, while setting the dimensional regulator to numerical values and solving the linear systems of equations appearing in the intermediate stages with floating-point arithmetic rather than finite fields.

The rest of this article is organised as follows. 
In \cref{sec:amplitude}, we establish our conventions, discuss the general structure of the UV and IR singularities, and define the two-loop hard function entering the NNLO cross section. 
In \cref{sec:thv}, we present the computation performed in the 't~Hooft-Veltman scheme; in particular, we describe the generation of the integrand and the definition of the scalar form factors (\cref{sec:integrand_generation}), as well as the reduction to special functions and the determination of the corresponding rational coefficients (\cref{sec:amplitude_coefficients}).
In \cref{sec:special_functions}, we discuss the construction of the set of special functions to represent the MIs.
In \cref{sec:numerical_eval}, we explain how the two-loop finite remainder is assembled and evaluated on a five-dimensional grid of phase-space points, and we present numerical results for several benchmark points together with the checks that we performed.
In \cref{sec:cdr}, we discuss an independent computation performed in the conventional dimensional regularisation scheme using a significantly different approach. This provides a stringent check of our results.
We draw our conclusions in \cref{sec:conclusions}.
Finally, we include two appendices.
In Appendix \ref{sec:appendixA}, we collect the explicit expressions for the renormalisation constants and anomalous dimensions governing the UV and IR singularities of the two-loop amplitude, while Appendix \ref{app:appendixB} presents the basis of Lorentz tensors used in its decomposition.

\section{Conventions, amplitude decomposition and pole structure}
\label{sec:amplitude}

In this section, we introduce our notation for the kinematics, define the leading-colour approximation of the amplitude, and discuss the structure of UV and IR poles.

%%%%%%%%%%%%%%%%%%%%%%%%%%%%%%
\subsection{Kinematics and conventions}
\label{sec:conventions}

We compute the two-loop QCD corrections to the scattering amplitude for the partonic process
\begin{equation}
\label{eq:ScatteringProcess}
	\overline{u}(p_1) + d(p_2) + \overline{t}(p_3) + t(p_4) + W^+(p_5) \to 0 \,,
\end{equation}
where an on-shell top--antitop quark pair is produced in association with a $W$ boson.
We assume the CKM matrix to be diagonal, thus the process can only be initiated by light quarks of different flavours, which we denote by $u$ and $d$.
Since the coupling of the $W$ boson to light quarks is flavour-independent, 
the scattering amplitude for the process in \cref{eq:ScatteringProcess} is equal to that for the following process:
\begin{equation}
\label{eq:ScatteringProcess2}
	\overline{d}(p_1)+u(p_2)+\overline{t}(p_3)+t(p_4)+W^-(p_5)\rightarrow 0\,.
\end{equation}
All external momenta $p_i$ in \cref{eq:ScatteringProcess} are taken to be incoming, and satisfy momentum conservation
\begin{equation}
	\sum_{i=1}^{5}p_i=0 \,,
\end{equation}
and on-shell conditions
\begin{equation}
	p_1^2=p_2^2=0 \,,  \qquad \ p_3^2=p_4^2=m_t^2 \,,  \qquad \ p_5^2=m_W^2 \,,
\end{equation}
where $m_t$ and $m_W$ denote the top-quark and $W$-boson mass, respectively.
A complete description of the process kinematics requires seven independent Lorentz invariants, 
\begin{equation}
\label{eq:invariants}
	\vec{x} =\bigl( s_{13},\,s_{34},\,s_{24},\,s_{25},\,s_{15},\,m_t^2, \,m_W^2\bigr) \, ,
\end{equation}
chosen according to the conventions of Ref.~\cite{Becchetti:2025osw} with $s_{ij}=(p_i+p_j)^2$,
and a pseudo-scalar invariant, 
\begin{equation}
\label{eq:tr5}
	\mathrm{tr}_5 = \mathrm{tr}\left(\gamma_5 \slashed{p}_1 \slashed{p}_2 \slashed{p}_3 \slashed{p}_4 \right)\,,
\end{equation}
whose sign encodes the parity-odd degree of freedom.
The square of $\mathrm{tr}_5$ is related to the scalar invariants $\vec{x}$ via
\begin{equation}
\label{eq:tr5_to_invariants}
	\mathrm{tr}_5^2=\mathbb{G}(p_1,p_2,p_3,p_4) \, ,
\end{equation}
where $\mathbb{G}$ denotes the Gram determinant of the Lorentz vectors in the argument, defined by
\begin{equation}
\label{eq:gram-definitions}
\mathbb{G}(a_1,\ldots,a_n )  = \det\begin{pmatrix}
2\, a_1 \cdot a_1 & \cdots & 2\, a_1\cdot a_n \\
\vdots & \ddots & \vdots \\
2\, a_n \cdot a_1 & \cdots & 2\, a_n\cdot a_n
\end{pmatrix}.
\end{equation}
In the notation of \cref{eq:ScatteringProcess}, the physical region corresponding to $\ttW$ production is the $12 \to 345$ channel, which is characterised by the following inequalities for the external momenta,
\begin{equation}
\label{eq:physical_region_1}
\begin{aligned}
&p_1 \cdot p_2 > 0\,, \quad p_1 \cdot p_3 < 0\,, \quad  p_1 \cdot p_4 <0 \,, \quad p_1 \cdot p_5 <0 \,, \quad  p_2 \cdot p_3 <0 \,,  \\
& p_2 \cdot p_4 <0 \,, \quad p_2 \cdot p_5 < 0 \,, \quad p_3 \cdot p_4 > 0 \,, \quad  p_3\cdot p_5 > 0 \,, \quad p_4\cdot p_5 > 0\,,
\end{aligned}
\end{equation}
together with the Gram-determinant constraints,
\begin{equation}
\label{eq:physical_region_2}
 \mathbb{G}(p_i, p_j) < 0 \,, \qquad \mathbb{G}(p_i,p_j,p_k) > 0 \,, \qquad \mathbb{G}(p_1,p_2,p_3,p_4) < 0 \,,
\end{equation}
for all distinct $i, j, k \in \{1,\ldots,5\}$.

While we treat the loop momenta in $d = 4 - 2 \eps$ space-time dimensions, we take the external momenta and polarisation states to be either four-dimensional in the 't Hooft-Veltman scheme (tHV, see \cref{sec:thv}) or $d$-dimensional in conventional dimensional regularisation (CDR, see \cref{sec:cdr}).

We adopt the \textit{generalised leading colour approximation} (LCA): the number of colours, $N_c=3$, and the number of light quark flavours, $n_l=5$, are treated as parametrically large, while contributions suppressed by inverse powers of $N_c$ or $n_l$ are neglected.
As a consequence, non-planar diagrams and diagrams involving closed top-quark loops do not contribute.
In this approximation, we write the $\ell$-loop bare scattering amplitude $\M^{(\ell)}$ for the process in \cref{eq:ScatteringProcess}~as 
\begin{equation}
\label{eq: bare_amplitude}
	\M^{(\ell)} = \frac{g_w}{\sqrt{2}} (4 \pi \asbare) \left[ (4\pi e^{-\gamma_E} )^\eps \,\frac{\asbare}{4 \pi} \right]^{\ell} \delta_{\ i_4}^{\bar{i}_1} \, \delta_{\ i_2}^{\bar{i}_3} \A^{(\ell)} \, ,
\end{equation}
where $\asbare=(g_s^0)^2/(4\pi)$, $g_s^0$ is the (dimensionful) bare strong coupling constant, $g_w$ is the weak coupling constant, and $\delta^{\bar{i}}_{\ j}$ is the Kronecker delta in the fundamental representation of $\mathrm{SU}(N_c)$, with $i_k$ ($\bar{i}_k$) denoting the colour index of the quark (antiquark) carrying momentum $p_k$.
In the LCA, the $\ell$-loop partial amplitudes $\A^{(\ell)}$ in \cref{eq: bare_amplitude} are decomposed according to their $N_c$ and $n_l$ contributions up to two-loop order as
\begin{subequations}
\label{eq: colour-decomposition}
\begin{align}
	\A^{(0)} &= A^{(0)}_{ \{0,0\} }  \,, \\
	\A^{(1)} &= N_c \,A^{(1)}_{ \{1,0\} } + n_l \,A^{(1)}_{ \{0,1\} } \,, \\
	\A^{(2)} &= N_c^2 \,A^{(2)}_{ \{2,0\} } + N_c\,  n_l \, A^{(2)}_{ \{1,1\} } + n_l^2 \, A^{(2)}_{ \{0,2\} } \,.
\end{align}
\end{subequations}
Throughout the remainder of this paper, we will often use $A^{(\ell)}$, omitting the subscript, to denote an arbitrary term in the  $(N_c,n_l)$ expansion given in \cref{eq: colour-decomposition}.

%%%%%%%%%%%%%%%%%%%%%%%%%%%%%%
\subsection{Ultraviolet and infrared pole structure}
\label{sec:polestructure}

To obtain UV and IR finite remainders, we first define the $\ell$-loop renormalised partial amplitudes as
\begin{subequations}
\label{eq:UVren_amplitudes}
\begin{align}
	 \A^{(1)}_{\mathrm{ren}} &=  \mu^{2 \eps} \A^{(1)} + \left(\delta Z_{\as}^{(1)} + \delta Z_{t}^{(1)} \right) \A^{(0)}  \, , \\
	 \A^{(2)}_{\mathrm{ren}}  &=  \mu^{4 \eps} \A^{(2)} + \mu^{2 \eps} \left(2 \, \delta Z_{\as}^{(1)} + \delta Z_{t}^{(1)} \right)   \A^{(1)}  
	 + \mu^{2 \eps} \delta Z_{m}^{(1)}  \A^{(1)}_{\mathrm{mct}} 
	 + \left(\delta Z_{\as}^{(2)} + \delta Z_{\as}^{(1)} \delta Z_{t}^{(1)} + \delta Z_{t}^{(2)} \right) \A^{(0)}  \,,
\end{align}
\end{subequations}
where $\A^{(1)}_{\mathrm{mct}}$ stands for the one-loop partial amplitude with the insertion of a single mass counterterm and $\mu$ is the renormalisation scale. 
The superscript in the renormalised quantities, such as $\A^{(\ell)}_{\mathrm{ren}}$ in \cref{eq:UVren_amplitudes}, refers to the order in the renormalised coupling $\as(\mu)/(4 \pi)$, defined via 
\begin{equation}
	\asbare \, (4\pi e^{-\gamma_E} )^\eps = \as(\mu) \, \mu^{2\eps} Z_{\as}\,,
\end{equation}
where
\begin{equation}
\label{eq: coupling_ren_constant}
	Z_{\as} = 1 + \frac{\as(\mu)}{4 \pi} \delta Z_{\as}^{(1)} +  \left(\frac{\as(\mu)}{4 \pi}\right)^2 \delta Z_{\as}^{(2)} + \mathcal{O}(\as^3) 
\end{equation}
is the coupling renormalisation constant in the $\MSbar$ scheme with $n_l$ active flavours.
The explicit expressions of $\delta Z_{\as}^{(\ell)}$, as well as of the counterterms $\delta Z_{m}^{(\ell)}$ and $\delta Z_{t}^{(\ell)}$ originating from the renormalisation of the top-quark mass and wavefunction, are provided in Appendix~\ref{sec:appendixA}.

Once the UV singularities are removed, the renormalised partial amplitudes $\A^{(\ell)}_{\mathrm{ren}}$ are still affected by divergences of IR origin, whose structure is completely understood at the two-loop level \cite{Catani:1998bh,Gardi:2009qi,Gardi:2009zv,Becher:2009cu,Becher:2009qa,Becher:2009kw,Ferroglia:2009ep,Ferroglia:2009ii}.
We therefore define the \textit{finite remainder} for each partial amplitude in the $\MSbar$ scheme~\cite{Becher:2009cu,Becher:2009qa,Becher:2009kw,Ferroglia:2009ep,Ferroglia:2009ii} as
\begin{subequations}
\label{eq: partial_finrem}
\begin{align}
	 \F^{(0)} &= \A^{(0)} \,, \\
	 \F^{(1)} &= \A^{(1)}_{\mathrm{ren}}  - \mathbf{Z}^{(1)}\A^{(0)} \, , \\
	 \F^{(2)} &= \A^{(2)}_{\mathrm{ren}}  - \mathbf{Z}^{(1)}\A^{(1)}_{\mathrm{ren}} - \left[ \mathbf{Z}^{(2)} - \left(\mathbf{Z}^{(1)}\right)^2\right]\A^{(0)}\,,
\end{align}
\end{subequations}
where
\begin{subequations}
\begin{align}
\mathbf{Z}^{(1)} & = \frac{\Gamma^{(1)'}}{4 \eps^2} + \frac{\mathbf{\Gamma}^{(1)}}{2 \eps}  \,, \\
\mathbf{Z}^{(2)} & = \frac{(\Gamma^{(1)'})^2}{32 \eps^4} + \frac{\Gamma^{(1)'}}{8 \eps^3}\left(\mathbf{\Gamma}^{(1)} -\frac{3}{2}\beta_0 \right) + \frac{\mathbf{\Gamma}^{(1)}}{8 \eps^2}\left(\mathbf{\Gamma}^{(1)} -2\beta_0 \right) + \frac{\Gamma^{(2)'}}{16 \eps^2} + \frac{\mathbf{\Gamma}^{(2)}}{4 \eps} \,,
\end{align}
\end{subequations}
with
\begin{align}
\Gamma^{(\ell)'} = \mu \, \frac{\partial \mathbf{\Gamma}^{(\ell)}}{\partial \mu} \,.
\end{align}
The perturbative coefficients of the anomalous dimension, $\mathbf{\Gamma}^{(\ell)}$, are generally operators in colour space and, for processes involving massive partons, can be found in Refs.~\cite{Ferroglia:2009ep,Ferroglia:2009ii}. 
For the process in \cref{eq:ScatteringProcess} in the LCA, they reduce to complex numbers, given by
\begin{subequations}
\label{eq: anomalous_dimension}
\begin{align}
	\Gamma^{(\ell)'} &= - N_c \, \gamma_{\mathrm{cusp}}^{(\ell)}  \,,\\
	\mathbf{\Gamma}^{(\ell)} &= 2 \left(\gamma_q^{(\ell)} + \gamma_Q^{(\ell)} \right) - \frac{N_c}{2} \, \gamma_{\mathrm{cusp}}^{(\ell)} \left[ \log\left(\frac{\mu^2}{m_t^2}\right) - \log\left(\frac{m_t^2-s_{14}}{m_t^2}\right) - \log\left(\frac{m_t^2-s_{23}}{m_t^2}\right) \right] \,,
\end{align}
\end{subequations}
where $s_{23}$ and $s_{14}$ can be expressed in terms of the seven independent Lorentz invariants given in \cref{eq:invariants}.
The perturbative coefficients of the cusp and collinear anomalous dimensions, $\gamma_{\mathrm{cusp}}^{(\ell)}$ and $\gamma_{q/Q}^{(\ell)}$ respectively, are provided in Appendix~\ref{sec:appendixA}.
The partial finite remainders $\F^{(\ell)}$, as well as the partial renormalised amplitudes $\A^{(\ell)}_{\mathrm{ren}}$, admit a decomposition in $N_c$ and $n_l$ contributions analogous to \cref{eq: colour-decomposition} for the bare partial amplitudes.

We use the partial finite remainders $\F^{(\ell)}$ to define the colour- and polarisation-summed \textit{hard function} $\Hard$ contributing to the cross section,
\begin{subequations}
\label{eq: hard-function_definition}
\begin{align}
	\Hard^{(0)} &= \overline{\sum_{\text{col}}}\, \overline{\sum_{\text{pol}}}\bigl| \R^{(0)} \bigr|^2 \,, \\
	\Hard^{(\ell)} &= 2 \, \text{Re} \, \overline{\sum_{\text{col}}}\, \overline{\sum_{\text{pol}}} \left(\R^{(0)}\right)^{\!*} \!\R^{(\ell)} \,,
\end{align}
\end{subequations}
where we sum over colour and polarisation of all external partons, with the overlines indicating the average for the initial-state quarks,
and $\R^{(\ell)}$ is the $\ell$-loop colour-dressed finite remainder.
In the LCA, the latter is given in terms of the partial finite remainder by 
\begin{equation}
\label{eq: colour-dressed_finrem}
	\R^{(\ell)} = \frac{g_w}{\sqrt{2}} \bigl(4 \pi \as(\mu)\bigr) \left( \frac{\as(\mu)}{4 \pi} \right)^{\ell} \delta_{\ i_4}^{\bar{i}_1} \, \delta_{\ i_2}^{\bar{i}_3} \F^{(\ell)} \,,
\end{equation}
while the perturbative coefficients of the hard function up to two-loop order\footnote{Note that the contribution from the one-loop squared finite remainder $\bigl| \R^{(1)} \bigr|^2$ is not included in our definition of the two-loop hard function $\Hard^{(2)}$.} read
\begin{subequations}
\label{eq: hard-function}
\begin{align}
	\Hard^{(0)} & =  \frac{N_c^2}{36} \frac{g_w^2}{2} \bigl(4 \pi \as(\mu) \bigr)^2 \, \sum_{\text{pol}}\bigl| \A^{(0)} \bigr|^2 \,, \\
	\Hard^{(1)} &=  \frac{N_c^2}{18} \frac{g_w^2}{2} \bigl(4 \pi \as(\mu) \bigr)^2 \left(\frac{\as(\mu)}{4 \pi}\right) \text{Re} \sum_{\text{pol}} \left(\F^{(0)}\right)^{\!*} \!\F^{(1)}  \,,\\
	\Hard^{(2)} & = \frac{N_c^2}{18} \frac{g_w^2}{2} \bigl(4 \pi \as(\mu)\bigr)^2 \left(\frac{\as(\mu)}{4 \pi}\right)^{\! 2} \text{Re} \sum_{\text{pol}} \left(\F^{(0)}\right)^{\!*} \!\F^{(2)}  \,.
\end{align}
\end{subequations}

\section{Calculation in the 't Hooft-Veltman scheme}
\label{sec:thv}

In this section, we discuss the calculation of the two-loop finite remainder for the process in \cref{eq:ScatteringProcess} in the 't~Hooft-Veltman scheme~\cite{tHooft:1972tcz}, in which the loop momenta and internal states are treated in $d$ dimensions, while the external momenta and polarisation states remain four-dimensional.
We begin with the generation of the bare amplitude from Feynman diagrams and its decomposition into form factors expressed in terms of scalar loop integrals.
We then discuss the reduction of the resulting integrands onto the set of special functions constructed in \cref{sec:special_functions}, which enables the extraction of the finite remainders.

%%%%%%%%%%%%%%%%%%%%%%%
\subsection{Integrand generation}
\label{sec:integrand_generation}

We decompose the coefficients of the $(N_c,n_l)$ expansion of the $\ell$-loop bare partial amplitudes in \cref{eq: colour-decomposition} into a basis of $N$ Lorentz tensor structures, $T_i$, multiplied by scalar form factors, $F_{\{a,b\}, i}^{(\ell)}$, as
\begin{equation}
\label{eq:tensordecomposition}
	A^{(\ell)}_{\{a,b\}} = \sum_{i=1}^N F_{\{a,b\}, i}^{(\ell)} \, T_i \,.
\end{equation} 
For simplicity, we hereafter suppress the subscript $\{a,b\}$ in $A^{(\ell)}$ and $F_{i}^{(\ell)}$ referring to the powers of $N_c$ and $n_l$ in \cref{eq: colour-decomposition}, as the decomposition applies independently to each of them.
In order to benefit from the simplifications that arise from considering four-dimensional external states in the tHV scheme, we employ the method of physical projectors to define the tensor basis~\cite{Peraro:2019cjj,Peraro:2020sfm}.
In this method, the tensor basis is independent of the loop order and its dimension is bounded from above by the number of independent  external polarisation states.
Since the light-quark current has two independent helicity configurations, the two top quarks admit two spin configurations each, while the $W$ boson has three physical polarisation states, the tensor basis contains $N=2 \times 2^2 \times 3=24$ elements.
We employ the tensor basis used at one loop in Ref.~\cite{Becchetti:2025osw}, multiplied by suitable factors of $m_t$ to ensure that all tensors have the same mass dimension.
We refer the reader to Ref.~\cite{Becchetti:2025osw} for a detailed discussion of the construction of this tensor basis, and give the explicit expressions for the tensors $T_i$ in Appendix~\ref{app:appendixB}.
The form factors may then be obtained by applying the projector operators,
\begin{equation}
\label{eq:def_projector}
P_i=\sum_{j=1}^N \bigl(M^{-1}\bigr)_{ij} \, T^\dagger_j \,,
\end{equation}
to the partial amplitudes $A^{(\ell)}$.
The matrix $M_{ij}$ in \cref{eq:def_projector} is defined as
\begin{equation}
\label{eq:defM}
M_{ij} = \sum_{\mathrm{pol}} T^\dagger_i T_j \,,
\end{equation}
where the sum runs over the spins and polarisations of the external particles. 
However, it is algebraically simpler to project the bare partial amplitudes onto the individual tensor structures $T_i$, defining the bare \textit{projected amplitudes} $B^{(\ell)}_i$,
\begin{equation}
\label{eq:Bs}
B^{(\ell)}_i = \sum_{\mathrm{pol}} T_i^\dagger\,A^{(\ell)}\,.
\end{equation}
As for $A^{(\ell)}$, we omit the subscripts of $B^{(\ell)}_i$ referring to the $(N_c,n_l)$ decomposition in \cref{eq: colour-decomposition}.
Once the projected amplitudes have been computed, the corresponding form factors are obtained via
\begin{equation}
\label{eq:FFfromB}
F_i^{(\ell)}=\sum_{j=1}^N \bigl(M^{-1}\bigr)_{ij} \, B^{(\ell)}_j\,.
\end{equation}
Alternatively, one can use the projected amplitudes to directly construct the interference between the loop and the tree-level bare partial amplitudes, as
\begin{equation}
\label{eq:interference}
\sum_{\text{pol}}\bigl(A^{(0)}\bigr)^* A^{(\ell)} = \bigl(B_i^{(0)}\bigr)^* \bigl(M^{-1}\bigr)^{\dagger}_{ij} \, m_{jk} \, \bigl(M^{-1}\bigr)_{kl} \, B_l^{(\ell)} \, ,
\end{equation}
where, for brevity, summation over repeated indices ($i,j,k,l=1,\ldots,24$) is understood.
The matrix $m_{ij}$ is defined analogously to $M_{ij}$ in \cref{eq:defM}, with the massless spinors in the basis tensors replaced as
\begin{equation}
m_{ij}= \sum_{\mathrm{pol}} \left. T^\dagger_i T_j \right|_{\substack{ {\overline{v}_1 \to \overline{v}_1^L} \\ {u_2 \to u_2^L}}}\, ,
\end{equation}
with $u_2^L = P_L\,u_2$ and $\overline{v}_1^{L}=\overline{v}_1 P_R$, where $P_{R/L}=(\mathds{1}\pm\gamma_5)/2$ are the chiral projectors.
These projectors account for the fact that the $W$ boson couples only to the left-handed massless quark current.
As a consequence, the matrix $m_{ij}$ depends linearly on the parity-odd invariant $\mathrm{tr}_5$ defined in \cref{eq:tr5}, as
\begin{equation} \label{eq:mijtr5}
m_{ij}= m_{ij}^{(1)} + \mathrm{tr}_5 \, m_{ij}^{(\mathrm{tr}_5)} \, .
\end{equation}
We emphasise that the projected amplitudes $B^{(\ell)}_i$ are scalar quantities, and the parity degree of freedom of the hard functions is entirely captured by $\mathrm{tr}_5$ in \cref{eq:mijtr5}.
Moreover, note that the last six projected amplitudes~---~$B^{(\ell)}_i$ with $i=19,\ldots,24$~---~drop out of the interference with the tree-level amplitude in \cref{eq:interference}, and thus do not contribute to the unpolarised hard functions defined in \cref{eq: hard-function}.
Nonetheless, we set up the computation of all projected amplitudes in order to retain the complete information about the polarisations of the external states.

\begin{figure}
\centering
\subfloat[Family ${\rm F}_1$]{\includegraphics[scale=0.3]{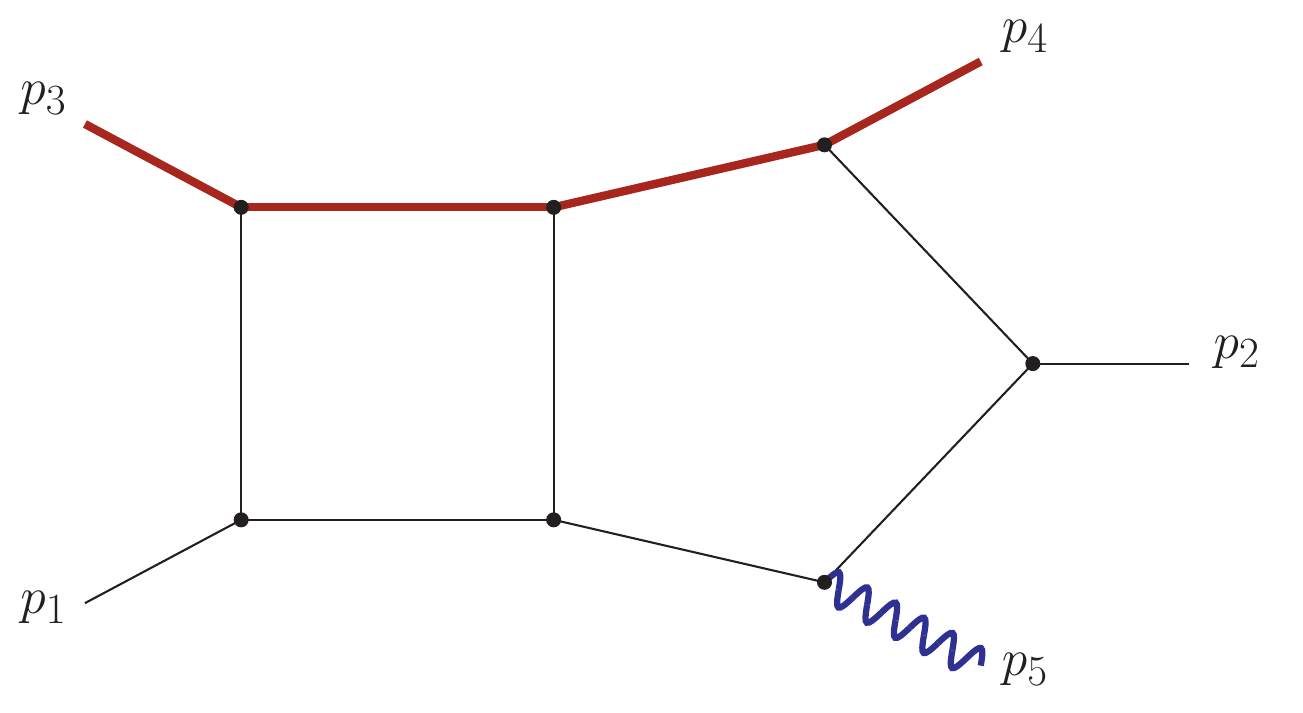}\label{fig:F1}}\quad
\subfloat[Family ${\rm F}_2$]{\includegraphics[scale=0.3]{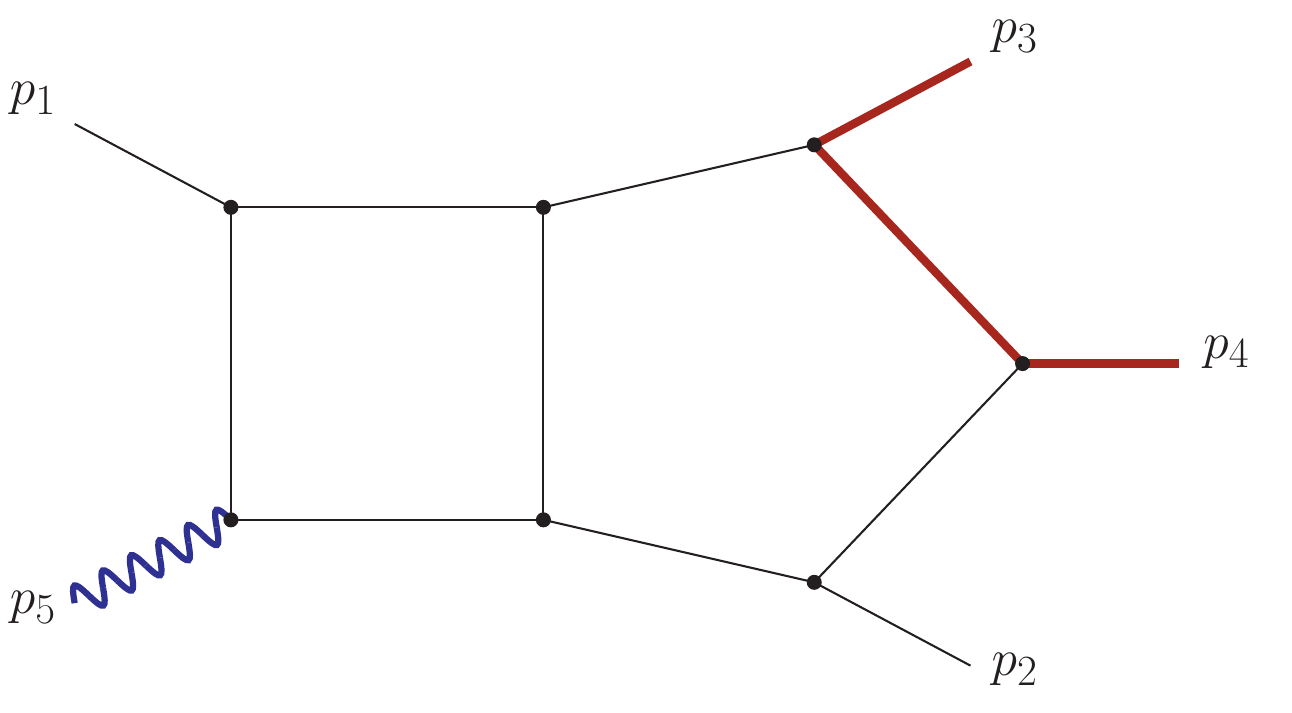}\label{fig:F2}}\\
\subfloat[Family ${\rm F}_3$]{\includegraphics[scale=0.3]{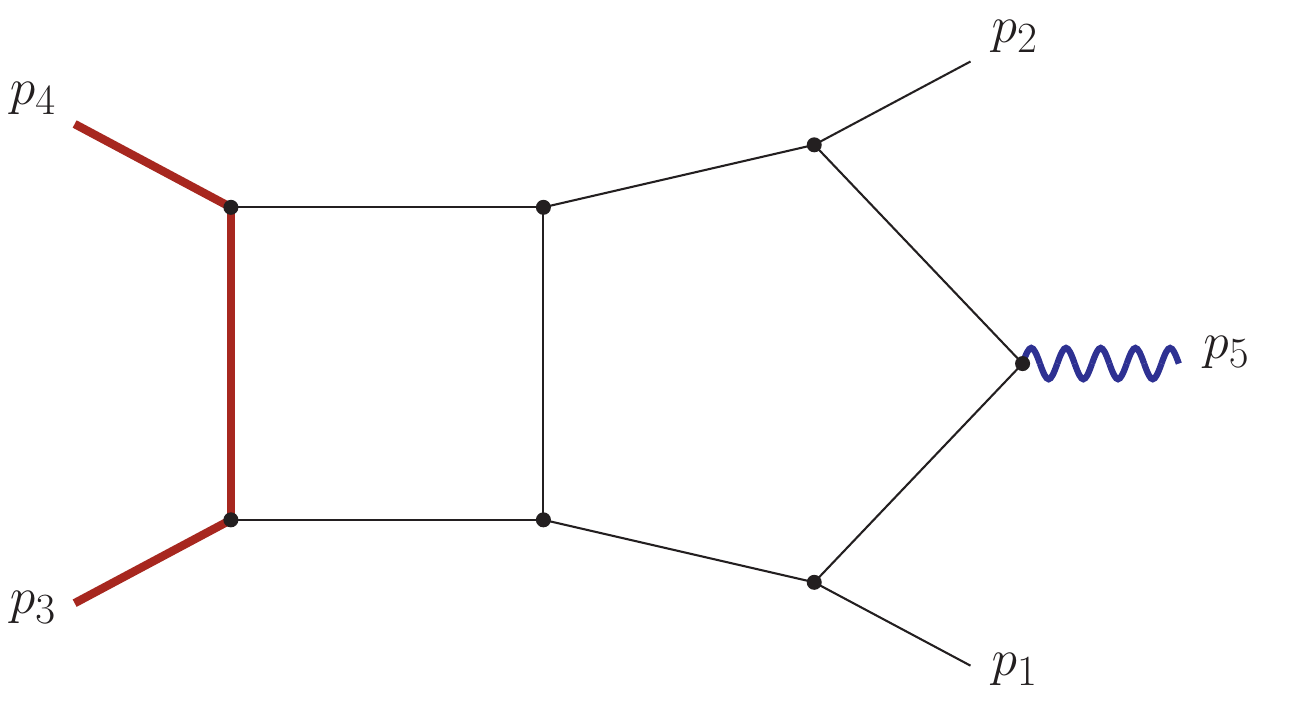}\label{fig:F3}}
\caption{Top-sector graphs of the irreducible two-loop families ${\rm F}_1$, ${\rm F}_2$ and ${\rm F}_3$. Thin black lines correspond to massless particles, thick red lines to (anti-)top quarks, and the curly blue line to the $W$ boson.}
\label{fig:top_sector_irreducible}
\end{figure}

\begin{table}[t!]
    \centering
    \begin{tabular}{c|c|c|c}
       & ${\rm F}_1$ & ${\rm F}_2$ & ${\rm F}_3$ \\
    \hline 
    $D_1$ & $k_1^2-m_t^2$ & $k_1^2$  & $k_1^2$ \\
    $D_2$ & $(k_1-p_3)^2$ & $(k_1-p_4)^2-m_t^2$  & $(k_1-p_2)^2$ \\
    $D_3$ & $(k_1-p_{23})^2$ & $(k_1-p_{34})^2$  & $(k_1-p_{25})^2$ \\
    $D_4$ & $(k_1-p_{235})^2$ & $(k_1-p_{234})^2$  & $(k_1+p_{34})^2$ \\
    $D_5$ & $k_2^2-m_t^2$ & $k_2^2$  & $k_2^2$ \\
    $D_6$ & $(k_2-p_4)^2$ & $(k_2+p_{2345})^2$  & $(k_2-p_3)^2-m_t^2$ \\
    $D_7$ & $(k_2+p_{235})^2$ & $(k_2+p_{234})^2$  & $(k_2-p_{34})^2$ \\
    $D_8$ & $(k_1+k_2)^2$ & $(k_1+k_2)^2$  & $(k_1+k_2)^2$ \\
    $D_9$ & $(k_1+p_4)^2-m_t^2$ & $(k_1-p_{2345})^2$  & $(k_1+p_3)^2-m_t^2$ \\
    $D_{10}$ & $(k_2+p_3)^2-m_t^2$ & $(k_2+p_4)^2-m_t^2$  & $(k_2+p_2)^2$ \\
    $D_{11}$ & $(k_2+p_{23})^2-m_t^2$ & $(k_2+p_{34})^2$  & $(k_2+p_{25})^2$ \\
\end{tabular}
    \caption{Definition of the inverse propagators for the two-loop irreducible integral families.
     We use the shorthand $p_{i_1\cdots i_k}$ defined in \cref{eq:pijk}.}
    \label{tab:Intfam1}
\end{table}

\begin{figure}
\centering
\subfloat[Family ${\rm F}_4$]{\includegraphics[scale=0.3]{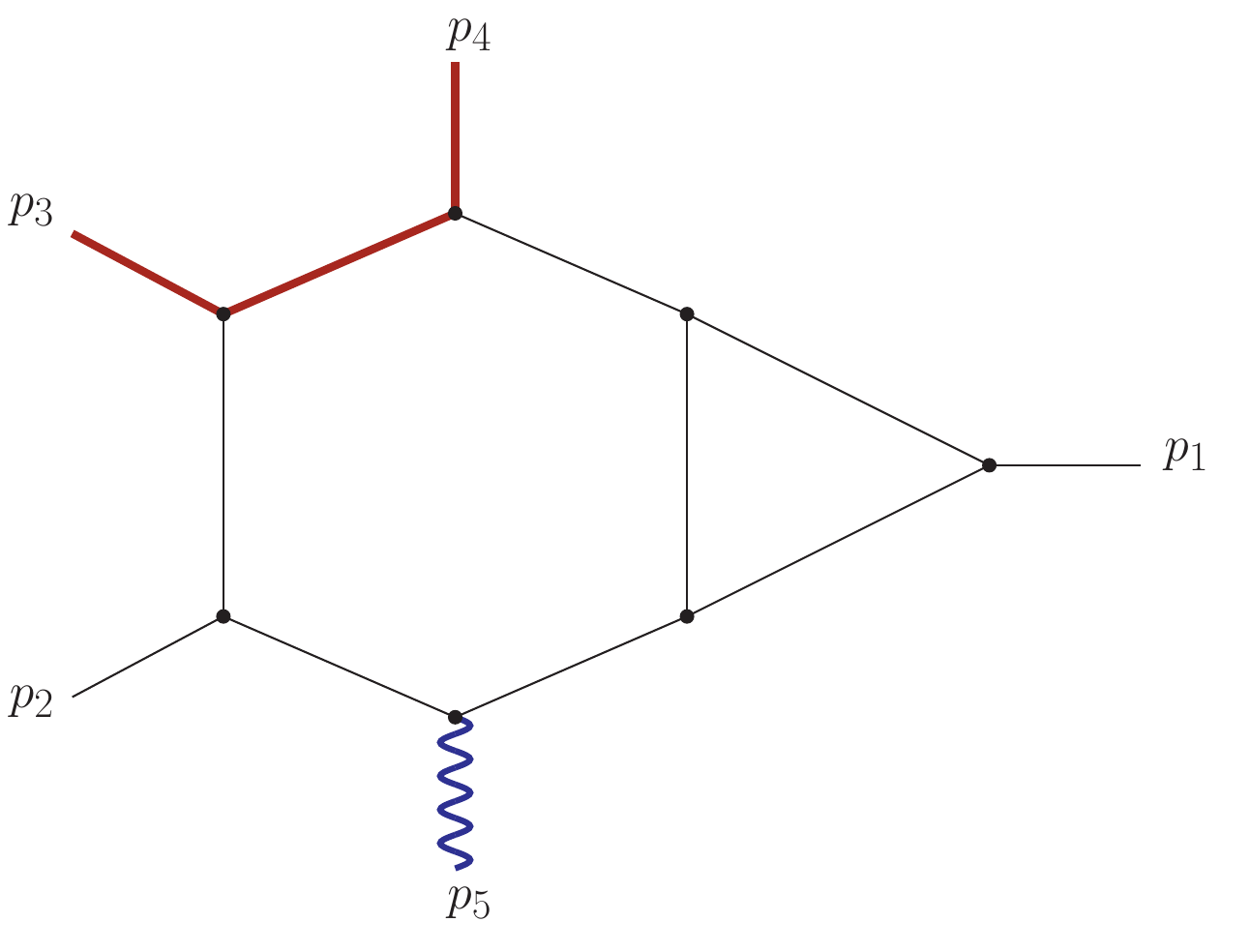}\label{fig:F4}}\quad
\subfloat[Family ${\rm F}_5$]{\includegraphics[scale=0.3]{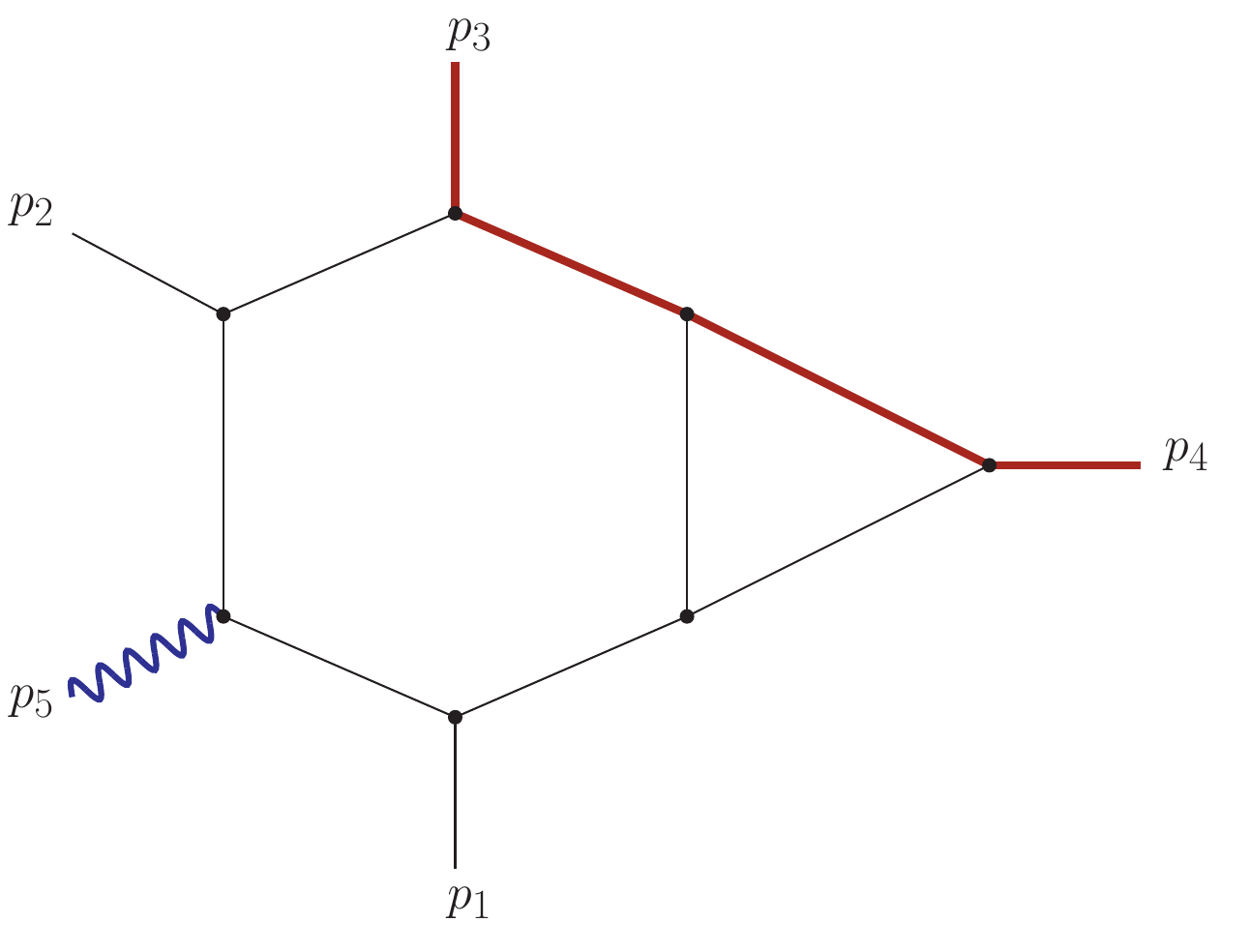}\label{fig:F5}}\\
\subfloat[Family ${\rm F}_6$]{\includegraphics[scale=0.3]{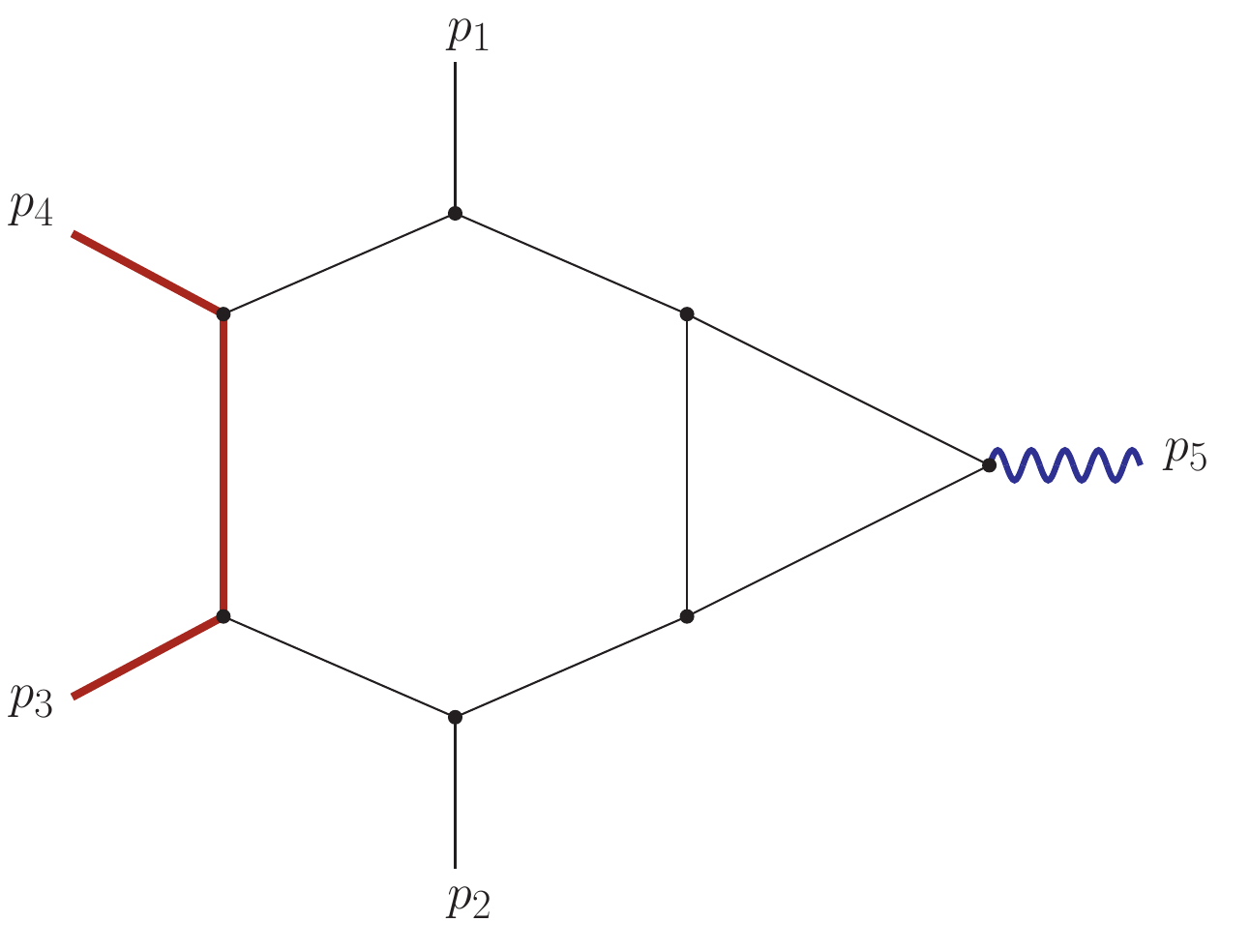}\label{fig:F6}}\quad
\subfloat[Family ${\rm A}^{(\textrm{x12})}$]{\includegraphics[scale=0.3]{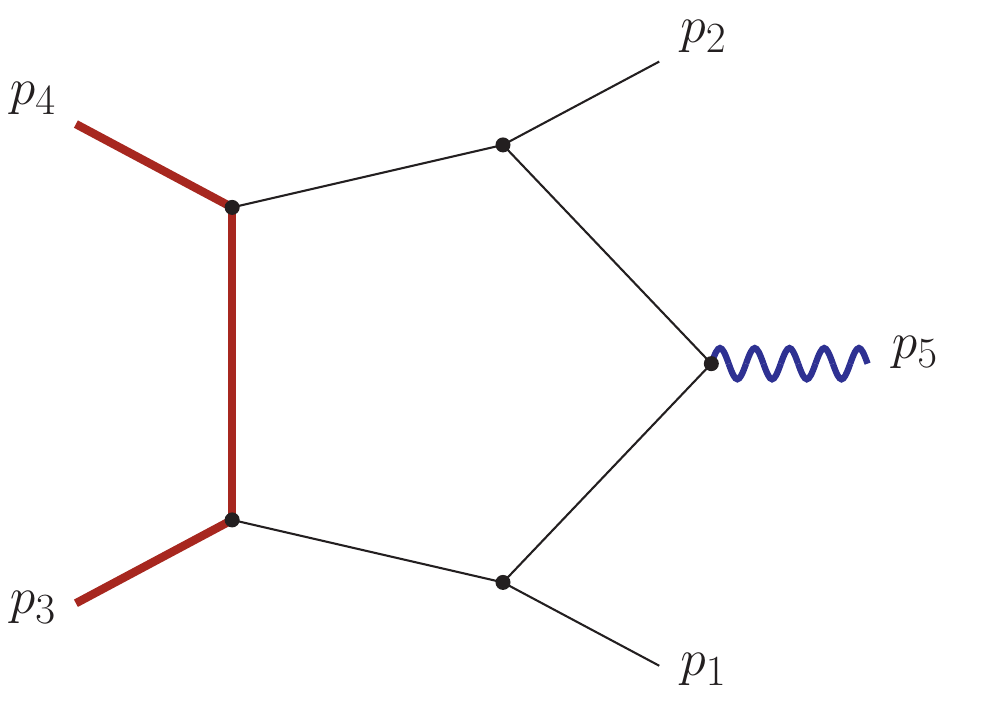}\label{fig:A}}
\caption{Top-sector graphs of the reducible two-loop families ${\rm F}_4$, ${\rm F}_5$ and ${\rm F}_6$, and of the one-loop family ${\rm A}^{(\textrm{x12})}$. Thin black lines correspond to massless particles, thick red lines to (anti-)top quarks, and the curly blue line to the $W$ boson.}
\label{fig:top_sector_reducible}
\end{figure}

In order to obtain an explicit integrand for the two-loop bare projected amplitudes,
we first generate all Feynman diagrams contributing to the process in \cref{eq:ScatteringProcess} using \textsc{QGRAF}~\cite{Nogueira:1991ex}. We then use \textsc{FORM}~\cite{Ruijl:2017dtg} to insert the Feynman rules and perform the colour algebra, thereby obtaining the colour decomposition of \cref{eq: colour-decomposition}. In total, 722 diagrams are generated, of which only 210 contribute to the LCA. For each of these 210 diagrams, we further use \textsc{FORM} to apply the projection given in \cref{eq:Bs} and carry out the resulting Dirac algebra separately. Performing this step on a diagram-by-diagram basis keeps the size of intermediate expressions manageable.
The projected diagrams can be expressed as a linear combination of scalar loop integrals with rational coefficients.
The loop integrals appearing in these expressions can be classified into six integral families, defined in \cref{tab:Intfam1,tab:Intfam2}, and their crossed counterparts.
In these tables, we use the shorthand,
\begin{equation} \label{eq:pijk}
 p_{i_1\cdots i_k}=\sum_{j\in\{i_1,\ldots,i_k\}} p_j \,.
\end{equation}
The required crossings correspond to the interchange of the light-quark momenta, $p_1\leftrightarrow p_2$ (denoted by $\textrm{x12}$), the top--antitop momenta, $p_3\leftrightarrow p_4$ (denoted by $\textrm{x34}$), and the simultaneous application of both crossings (denoted by $\textrm{x12x34}$).
More specifically, the following families appear:
\begin{equation}
\begin{split}
\{&{\rm F}_1, \, {\rm F}_2, \, {\rm F}_3, \, {\rm F}_4, \, {\rm F}_5, \, {\rm F}_6, \, {\rm F}_1^{\textrm{(x12)}}, \, {\rm F}_1^{\textrm{(x34)}}, \, {\rm F}_1^{\textrm{(x12x34)}}, \, {\rm F}_2^{\textrm{(x12)}}, \, {\rm F}_2^{\textrm{(x12x34)}}, \\
& {\rm F}_3^{\textrm{(x12)}}, \, {\rm F}_3^{\textrm{(x34)}}, \, {\rm F}_3^{\textrm{(x12x34)}}, \, {\rm F}_4^{\textrm{(x12)}}, \, {\rm F}_4^{\textrm{(x12x34)}},\,  {\rm F}_5^{\textrm{(x12x34)}}, \, {\rm F}_6^{\textrm{(x12)}}, \, {\rm F}_6^{\textrm{(x12x34)}} \} \, .
\end{split}
\end{equation}
We define the integrals of family ${\rm F} \in \{{\rm F}_1, {\rm F}_2, {\rm F}_3, {\rm F}_4, {\rm F}_5, {\rm F}_6 \}$ as
\begin{equation}
\label{eq:families}
I^{({\rm F})}_{\vec{\nu}} = \int \bigg(\prod_{l=1}^2 e^{\gamma_E \eps} \frac{\mathrm{d}^d k_l}{\ii\pi^{d/2}}\bigg) \prod_{j=1}^{11} \frac{1}{D^{\nu_j}_{{\rm F},j}} \,,
\end{equation}
where $\vec{\nu} = (\nu_1, \cdots, \nu_{11}) \in \mathbb{Z}^{11}$, and the inverse propagators $D_{{\rm F},j}$ are given in \cref{tab:Intfam1,tab:Intfam2}.
The last three inverse propagators~---~$D_{{\rm F},9}$, $D_{{\rm F},10}$ and $D_{{\rm F},11}$~---~are irreducible scalar products, i.e.\ $\nu_9, \nu_{10}, \nu_{11} \le 0$.
We follow the definitions of Ref.~\cite{Becchetti:2025qlu} for the families ${\rm F}_1$, ${\rm F}_2$ and ${\rm F}_3$.\footnote{To translate to the notation of the present work, the external momenta of Ref.~\cite{Becchetti:2025qlu} should be relabelled according to
$p_1\to p_4$, $p_2\to p_3$, $p_3\to p_2$, $p_4\to p_5$, and $p_5\to p_1$.}
The integrals belonging to the families ${\rm F}_4$, ${\rm F}_5$ and ${\rm F}_6$ can either be mapped onto integrals of the first three families, which were computed in Ref.~\cite{Becchetti:2025qlu}, or reduced to products of elements of the one-loop family ${\rm A}^{\textrm{(x12)}}$, which was studied in Ref.~\cite{Becchetti:2025osw}.\footnote{
We call the one-loop family ${\rm A}^{(\textrm{x12})}$ because it is obtained by exchanging $p_1$ and $p_2$ in the family ${\rm A}$ of Ref.~\cite{Becchetti:2025osw}. We use ${\rm A}^{(\textrm{x12})}$ rather than ${\rm A}$ in order to match the order of the external momenta of the two-loop families defined in \cref{tab:Intfam1,tab:Intfam2}.}
We define these one-loop integrals as
\begin{equation}
\label{eq:1L_families}
I^{({\rm F})}_{\vec{\nu}} = \int e^{\gamma_E \eps} \frac{\mathrm{d}^d k_1}{\ii\pi^{d/2}} \prod_{j=1}^{5} \frac{1}{D^{\nu_j}_{{\rm F},j}} \,,
\end{equation}
where the inverse propagators of the family ${\rm A}^{(\textrm{x12})}$ are given in \cref{tab:Intfam1L}. 
The top-sector graphs of all relevant families are shown in \cref{fig:top_sector_irreducible,fig:top_sector_reducible}.

\begin{table}[t!]
    \centering
    \begin{tabular}{c|c|c|c}
      & ${\rm F}_4$ & ${\rm F}_5$ & ${\rm F}_6$ \\
    \hline 
    $D_1$ & $k_1^2$ & $k_1^2$  & $k_1^2$ \\
    $D_2$ & $(k_1-p_4)^2-m_t^2$ & $(k_1-p_4)^2-m_t^2$  & $(k_1+p_5)^2$ \\
    $D_3$ & $(k_1-p_{34})^2$ & $(k_1-p_{34})^2$  & $(k_1+p_{25})^2$ \\
    $D_4$ & $(k_1-p_{234})^2$ & $(k_1-p_{234})^2$  & $(k_1+p_{235})^2-m_t^2$ \\
    $D_5$ & $(k_1-p_{2345})^2$ & $(k_1-p_{2345})^2$  & $(k_1+p_{2345})^2$ \\
    $D_6$ & $k_2^2$ & $k_2^2$  & $k_2^2$ \\
    $D_7$ & $(k_2+p_{2345})^2$ & $(k_2+p_{4})^2-m_t^2$  & $(k_2-p_{5})^2$ \\
    $D_8$ & $(k_1+k_2)^2$ & $(k_1+k_2)^2$  & $(k_1+k_2)^2$ \\
    $D_9$ & $(k_2+p_4)^2-m_t^2$ & $(k_2+p_{34})^2-m_t^2$  & $(k_2-p_{25})^2-m_t^2$ \\
    $D_{10}$ & $(k_2+p_{34})^2-m_t^2$ & $(k_2+p_{234})^2-m_t^2$  & $(k_2-p_{235})^2-m_t^2$ \\
    $D_{11}$ & $(k_2+p_{234})^2-m_t^2$ & $(k_2+p_{2345})^2-m_t^2$  & $(k_2-p_{2345})^2-m_t^2$ \\
\end{tabular}
    \caption{Definition of the inverse propagators for the two-loop reducible integral families. All integrals in these families are either reducible to integrals belonging to families ${\rm F}_1$, ${\rm F}_2$, ${\rm F}_3$, and their crossings, or to products of one-loop integrals for the same process. We use the shorthand $p_{i_1\cdots i_k}$ defined in \cref{eq:pijk}.}
    \label{tab:Intfam2}
\end{table}

With the classification of loop integrals at hand, we use \textsc{Reduze2}~\cite{vonManteuffel:2012np} to map the integrand to the unique sectors of these families.
We group diagrams by sectors, insert integral symmetries generated by \textsc{Reduze2}, simplify the rational coefficients with \textsc{FORM}, and only then sum over sectors to generate the full expression for each $B^{(2)}_i$.
Finally, the projected amplitudes can be expressed as linear combinations of scalar loop integrals, $\mathcal{I}_j$, with coefficients $c_{ij}$ that are rational functions of the kinematic invariants and dimensional regulator $\eps$,
\begin{equation}
\label{eq:scalar-decomposition}
	B^{(2)}_i(\vec{x}; \eps) = \sum_j c_{ij}(\vec{x};\eps) \, \mathcal{I}_j(\vec{x};\eps) \,.
\end{equation}
After including all crossings, the decomposition in \cref{eq:scalar-decomposition} involves 8959 scalar loop integrals.

\begin{table}[t!]
    \centering
    \begin{tabular}{c|c}
      & ${\rm A}^{(\textrm{x12})}$  \\
    \hline 
    $D_1$ & $k_1^2$ \\
    $D_2$ & $(k_1-p_{1345})^2$ \\
    $D_3$ & $(k_1-p_{145})^2 -m_t^2$\\
    $D_4$ & $(k_1-p_{15})^2$\\
    $D_5$ & $(k_1-p_{5})^2$\\
\end{tabular}
    \caption{Definition of the inverse propagators for the one-loop irreducible integral family. We use the shorthand $p_{i_1\cdots i_k}$ defined in \cref{eq:pijk}.}
    \label{tab:Intfam1L}
\end{table}

%%%%%%%%%%%%%%%%%%%%%%%
\subsection{Calculation of the finite remainder}
\label{sec:amplitude_coefficients}

We reduce the scalar integrals $\mathcal{I}_j$ appearing in \cref{eq:scalar-decomposition} to linearly independent MIs using integration-by-parts (IBP) relations~\cite{Tkachov:1981wb,Chetyrkin:1981qh,Laporta:2000dsw}.
These relations, as well as symmetry relations between integrals of the same family, have been generated using the \textsc{NeatIBP}~\cite{Wu:2023upw} program.\footnote{We used a customised version of \textsc{NeatIBP} which internally uses \textsc{FiniteFlow} as a linear solver, as we found it to be more stable for the integral families addressed in this work.}
We obtained additional symmetry relations between integrals of different families with the private \textsc{Mathematica} package~\textsc{FFIntRed}, which searches for shifts of the loop momenta that map sectors of one family onto sectors of another.
We collected these identities into a linear system of equations and solved it numerically over prime fields using \textsc{FiniteFlow}~\cite{Peraro:2019svx}. This solution expresses every integral contributing to \cref{eq:scalar-decomposition} as a linear combination
\begin{equation} \label{eq:preFFmis}
	B^{(2)}_i(\vec{x}; \eps) = \sum_{k=1}^{330} b_{ik}(\vec{x};\eps) \, {J}_k(\vec{x};\eps)
\end{equation}
of $330$ MIs denoted ${J}_k$, which belong to the families ${\rm F}_1$, ${\rm F}_1^{\textrm{(x12x34)}}$, ${\rm F}_2$, ${\rm F}_2^{\textrm{(x12)}}$, ${\rm F}_2^{\textrm{(x12x34)}}$, ${\rm F}_3$, ${\rm F}_5$ and ${\rm F}_6$.
The required MIs of the families ${\rm F}_5$ and ${\rm F}_6$ are given by linearly independent products of the MIs of family ${\rm A}^{\textrm{(x12)}}$ computed in Ref.~\cite{Becchetti:2025osw}. For the remaining families, we adopt the two-loop bases introduced in Ref.~\cite{Becchetti:2025qlu}. To ensure that the coefficients $b_{ik}$ are rational functions of the kinematic invariants, we remove the overall products of square roots from the definition of the MIs identified in that reference.

While the MIs are linearly independent in a generic number $d=4-2\eps$ of dimensions, additional relations exist among the coefficients of their expansion in the dimensional regulator $\eps$. 
Given that our ultimate goal is the calculation of the finite remainder (see \cref{eq: partial_finrem}), it is enough to expand \cref{eq:preFFmis} up to $\mathcal{O}(\eps^0)$.
To achieve this, we expressed the $\eps$-expansion of our MIs in terms of special functions, transcendental constants ($\zeta_2$ and $\zeta_3$), and square roots (henceforth collectively referred to as ``special functions'' for brevity), yielding an expansion for the bare projected amplitudes of the form
\begin{align} \label{eq:preFFexpr}
	B^{(2)}_i(\vec{x}; \eps) = \sum_{k=-4}^0  \sum_{j} \eps^k \, b_{ikj}(\vec{x}) \, Q_j(\vec{x}) + \mathcal{O}(\eps) \,,
\end{align}
where $b_{ikj}(\vec{x})$ are $\mathbb{Q}$-linearly independent rational coefficients, and $Q_j(\vec{x})$ are polynomials in the special functions, which will be discussed in \cref{sec:special_functions}.

In order to identify a basis of rational coefficients, we first write the expansion in terms of a set of monomials of special functions, and use numerical finite-field evaluations to find the $\mathbb{Q}$-linear relations among their rational coefficients (see Ref.~\cite{Badger:2021imn} for a thorough discussion of this procedure).
We then solve these relations to map the coefficients onto a basis, $\{b_{ikj}(\vec{x})\}$ in \cref{eq:preFFexpr}.
This basis is chosen to minimise the complexity, as measured by the total polynomial degree in $\vec{x}$ of the numerators, which is computed by reconstructing the coefficients on univariate slices.
When going from the initial set of 20406 linearly dependent coefficients to the 5780-dimensional basis, the maximal numerator (denominator) degree in $\vec{x}$ drops from 185 (182) to 157 (154).
While the overall complexity remains substantial, the simplification with respect to the coefficients $b_{ik}(\vec{x};\eps)$ of the MIs ${J}_k$ in \cref{eq:preFFmis} is significant. Indeed, the latter have maximal numerator and denominator degrees of 239 and 238, respectively, in the variables $\{\eps,\vec{x}\}$.
By contrast, Laurent expanding only the rational coefficients $b_{ik}(\vec{x};\eps)$ to the required $\eps$ order, without exploiting the relations between the coefficients of the expansion of the MIs themselves, would yield only a minor improvement, with the maximal numerator (denominator) degree in $\vec{x}$ reduced to 233 (230).
As we discuss below, even though we do not perform a full analytic reconstruction of the coefficients, this simplification is useful even in the context of a numerical calculation.

The $\eps$-poles present in \cref{eq:preFFexpr} are cancelled by adding the UV and IR subtraction terms defined in \cref{sec:polestructure} directly to the projected amplitudes, which obey the same renormalisation and subtraction relations as the corresponding partial amplitudes. In particular, we use \cref{eq:UVren_amplitudes} to construct the renormalised projected amplitudes $B_{\mathrm{ren},i}^{(\ell)}$ from the bare ones $B_i^{(\ell)}$. The mass counterterm entering the two-loop relation is obtained by applying the same projection to the one-loop amplitude with a single mass counterterm insertion. We then get the \textit{finite projected amplitudes} $B_{\mathrm{fin},i}^{(\ell)}$ from $B_{\mathrm{ren},i}^{(\ell)}$ using \cref{eq: partial_finrem}. We expand the resulting expressions in powers of $N_c$ and $n_l$ according to \cref{eq: colour-decomposition}, and express all projected amplitudes in terms of the same set of special functions.
To this end, we computed the leading-colour one-loop amplitude~---~with and without the insertion of the mass counterterm~---~up to order $\eps^2$ using the same setup employed at two loops.
The poles cancel out, and we obtain the finite projected amplitudes
\begin{equation} \label{eq:preFFfin}
B^{(2)}_{\textrm{fin},i}(\vec{x}) = \sum_{j} r_{ij}(\vec{x}) \, Q_j(\vec{x}) \,,
\end{equation}
necessary to construct the finite remainder in \cref{eq: partial_finrem}.
Once again, we are omitting from $B^{(2)}_{\textrm{fin},i}$ and $r_{ij}(\vec{x})$ the subscripts referring to the $(N_c,n_l)$ decomposition in \cref{eq: colour-decomposition}.
The rational coefficients $r_{ij}$ in \cref{eq:preFFfin} are $\mathbb{Q}$-linear combinations of the rational coefficients of the tree-level, one-loop, and two-loop projected amplitudes.
We stress that, at this stage, the polynomials $Q_j(\vec{x})$ are purely symbolic, hence the pole cancellation is exact.

The form factors of the tensor decomposition of the partial finite remainders $\mathcal{F}^{(\ell)}$ defined in \cref{eq: partial_finrem} may be obtained from the finite projected amplitudes $B^{(2)}_{\textrm{fin},i}$ by multiplying them by the inverse of the matrix $M$, as in \cref{eq:FFfromB} for the bare partial amplitudes, and summing all relevant $(N_c,n_l)$ terms.
In particular, it follows from \cref{eq:interference} that the finite-remainder interferences contributing to the hard functions in \cref{eq: hard-function} are expressed in terms of the finite projected amplitudes as
\begin{align} \label{eq:interferences_B}
 \sum_{\text{pol}} \bigl(\mathcal{F}^{(0)}\bigr)^* \mathcal{F}^{(\ell)} = \sum_{\substack{a,b=0 \\ a+b = \ell}}^{\ell} N_c^a \, n_l^b \, \Bigl(B^{(0)}_{\{0,0\}, i}\Bigr)^* \, \bigl( M^{-1}\bigr)^{\dagger}_{ij} \, m_{jk} \, \bigl( M^{-1}\bigr)_{kq} \, B^{(\ell)}_{\{a,b\},\textrm{fin},q} \,,
\end{align}
where we restored the subscript $\{a,b\}$ labelling the order in $N_c$ and $n_l$ in \cref{eq: colour-decomposition}, and repeated indices are summed over ($i,j,k,q=1,\ldots,24$).
The matrices $m$ and $M$, and the tree-level projected amplitudes $B^{(0)}_{\{0,0\}, i}$ are known analytically as functions of $\vec{x}$ and $\mathrm{tr}_5$.

This setup has been implemented using \textsc{FiniteFlow}, as extensively described in Ref.~\cite{Peraro:2019svx}, yielding an evaluation over finite fields of the coefficients $r_{ij}(\vec{x})$ of the finite projected amplitudes in \cref{eq:preFFfin}.
We process all projected amplitudes and all their $N_c$ and $n_l$ terms simultaneously.
At this stage, the coefficients $r_{ij}(\vec{x})$ could in principle be reconstructed analytically from finite-field evaluations. 
However, this would require a very large number of evaluations, and the resulting expressions would be very complicated, requiring considerable effort to simplify them into a usable form.
Moreover, since the evaluation of the special functions is the main bottleneck, compact analytic expressions for the rational coefficients would bring little practical benefit.
Therefore, we opted for the more pragmatic approach of evaluating such coefficients numerically over a grid covering the relevant phase space, which we used in Ref.~\cite{Becchetti:2026awn} to obtain the double virtual contribution to the NNLO QCD $\ttW$ cross section.

The numerical evaluation of the rational coefficients follows an approach proposed in Ref.~\cite{Peraro:2019okx}.
The points in the phase-space grid are rationalised with respect to the variables $\vec{x}$ (see \cref{sec:numerical_eval} for more details).
At each grid point, the coefficients are then evaluated over finite fields $\mathbb{Z}_p$ for several prime numbers $p$, and their exact rational values are subsequently reconstructed by combining the Chinese Remainder theorem and Wang's rational reconstruction algorithm~\cite{Wang:1981:PAU:800206.806398,Wang:1982:PRR:1089292.1089293}.
We use the \textsc{FiniteFlow} framework to easily parallelise both the numerical evaluations and the rational reconstruction.
The required evaluations are scheduled to maximise the usage of available computing threads and minimise prime switching (i.e., computing all required points for a prime before switching to the next prime) in each evaluation thread, as \textsc{FiniteFlow}'s algorithms typically cache prime-dependent and $\vec{x}$-independent parts of a calculation.

We stress that, although we do not perform an analytic reconstruction in this work, our approach of targeting the rational reconstruction of the coefficients $b_{ikj}(\vec{x})$ in \cref{eq:preFFexpr}, rather than the coefficients of the reduction to MIs, is still advantageous as it enables the analytic subtraction of the poles, and it requires significantly fewer primes and, consequently, fewer numerical evaluations. 
To quantify this improvement, we note that, for the phase-space point requiring the largest number of prime evaluations, reconstructing the linearly independent coefficients in \cref{eq:preFFexpr} requires \textit{only} 400 primes, compared with the 608 needed to reconstruct the $\eps$-expanded coefficients of the MIs.
More generally, we carried out this test at a few representative phase-space points, and observed a reduction of around $30\%$ in the number of needed prime fields.
We emphasise that the cancellations underlying this simplification occur exactly in the intermediate stages, thereby reducing the amount of numerical cancellations in the evaluation.
Further important benefits of this approach concern the numerical evaluation of special functions rather than the MIs, as will be discussed in the next section.

Although potentially less efficient than floating-point evaluation, finite fields offer several advantages. 
First, the calculation is exact, apart from the rationalisation of the phase-space point (which is, however, also required by the tools we use to evaluate the special functions).
As a result, the cancellation of the UV and IR poles is exact, and the finite remainder can be evaluated directly.
Second, this approach enables us to make extensive use of the routines already available in \textsc{FiniteFlow}.
Third, since our main bottleneck is the evaluation of the special functions, the performance penalty associated with these advantages is relatively small.
Finally, it allows us to check that the rational coefficients of the two-loop amplitude exhibit the expected singularities.

In fact, thanks to our choice of MIs, the denominators of the rational coefficients are expected to factor into $\eps$-dependent factors of the form $a \eps + b$ with $a,b \in \mathbb{Z}$~---~easily identified by reconstructing the dependence in $\eps$ for fixed $\vec{x}$~---~and polynomials in $\vec{x}$ drawn from the denominators of the DEs for the MIs of Ref.~\cite{Becchetti:2025qlu} (and permutations thereof)~\cite{Abreu:2018zmy}.
We can therefore determine the denominators analytically by matching an ansatz made of products of such factors against the analytic reconstruction on a univariate phase-space slice (see Ref.~\cite{Badger:2021imn} for a detailed discussion).
We find that the rational coefficients contain 61 irreducible polynomials in $\vec{x}$, all of which appear in the DEs for the MIs.
These include, in particular, the two highest-degree polynomials (of total degree 9 and 14) discussed in Ref.~\cite{Becchetti:2025qlu}.

\section{Master integrals and special functions}
\label{sec:special_functions}

In this section, we discuss our representation of the Laurent expansion of the MIs around $\eps=0$ in terms of special functions and transcendental constants. 
To this end, we employ the algorithm of Ref.~\cite{Badger:2024dxo}, which we briefly summarise here for completeness.

The algorithm takes as input the system of DEs for the set of MIs of each integral family, together with numerical evaluations of the MIs at a few phase-space points.
For the two-loop families ${\rm F}_1$, ${\rm F}_2$ and ${\rm F}_3$, we take MIs and DEs from Ref.~\cite{Becchetti:2025qlu}.
Permuting these appropriately yields MIs and DEs also for the permuted families (${\rm F}_1^{\textrm{(x12x34)}}$, ${\rm F}_2^{\textrm{(x12)}}$ and ${\rm F}_2^{\textrm{(x12x34)}}$).
Boundary values are obtained by means of \textsc{AMFlow}~\cite{Liu:2017jxz,Liu:2021wks,Liu:2022chg} to at least 35-digit precision.
The needed MIs of the families ${\rm F}_5$ and ${\rm F}_6$ factor into products of one-loop integrals computed 
in Ref.~\cite{Becchetti:2025osw}, which we recomputed in the same setup.
We stress that, in contrast to the previous section, here we adopt exactly the integral bases of the literature, i.e.\ we do not remove the square-root prefactors.

Denoting by $\vec{\mathcal{J}}_{\rm F}$ the set of MIs of the integral family ${\rm F}$, the DEs of Ref.~\cite{Becchetti:2025qlu} take the form
\begin{equation}
\dd \vec{\mathcal{J}}_{\rm F} (\vec{x}; \eps) = \left[\sum_{k=0}^2 \eps^k \dd {\rm M}^{({\rm F},k)} (\vec{x})\right] \cdot \vec{\mathcal{J}}_{\rm F} (\vec{x}; \eps) \,,
\label{eq:DEs}
\end{equation}
where $\dd$ is the total differential in the variables $\vec{x}$, and
\begin{align} \label{eq:dM}
 \dd {\rm M}^{({\rm F},k)}(\vec{x}) = \sum_{\alpha} a_{k \alpha}^{({\rm F})} \dd \log W_\alpha (\vec{x})  + \sum_{\beta} b_{k \beta}^{({\rm F})} \omega_{\beta}(\vec{x}) \,.
\end{align}
Here, $\omega_{\beta}(\vec{x})$ denotes a non-logarithmic one-form, which we write as
\begin{align} \label{eq:omega_expr}
\omega_{\beta}(\vec{x}) = \sum_{i=1}^7 \omega_{\beta i}(\vec{x}) \, \dd x_i \,.
\end{align}
The coefficients of the differentials $\omega_{\beta i}(\vec{x})$ in \cref{eq:omega_expr} as well as the arguments $W_\alpha(\vec{x})$ of the logarithmic one-forms in \cref{eq:dM}, called letters, are either rational functions or contain simple square roots.\footnote{Although the analytic structure of some of the MIs under consideration also involves nested square roots, these would explicitly appear in the DEs only if included in the definition of the MIs. We refrain from doing so, following Refs.~\cite{Badger:2024dxo,Becchetti:2025qlu}.}
We aim to expand the solution to \cref{eq:DEs} around $\eps=0$ up to the maximal order that is required to compute the two-loop finite remainder. 
In our case, we truncate the expansion as
\begin{equation}
\vec{\mathcal{J}}_{\rm F} (\vec{x}; \eps) = \sum_{w=0}^4 \eps^w \ \vec{\mathcal{J}}_{\rm F}^{(w)}(\vec{x})  + \mathcal{O}\left(\eps^5\right) \, .
\label{eq:MIs_Laurent_expansion}
\end{equation}
The coefficients $\vec{\mathcal{J}}_{\rm F}^{(w)}(\vec{x})$ satisfy algebraic relations. We therefore seek a set of algebraically independent special functions that simultaneously covers all relevant integral families.
This can be achieved algorithmically in the special case of \cref{eq:DEs} called canonical form~\cite{Henn:2013pwa}, where $\dd {\rm M}^{({\rm F},0)} = \dd {\rm M}^{({\rm F},2)} = 0$ and the solution can be written in terms of Chen iterated integrals~\cite{Chen:1977oja}.
This class of functions can be handled using well-established techniques that, in particular, enable the construction of a set of algebraically independent functions~\cite{Goncharov:2010jf}.
However, obtaining a canonical form for the two-loop families considered here is challenging, owing to the appearance of complicated analytic structures~---~including multiple elliptic curves and nested square roots~---~and to a high degree of algebraic complexity.

Instead, we follow the approach of Ref.~\cite{Badger:2024dxo}, which does not require a fully canonical form of the DEs.
The key idea is that, while the MIs $\vec{\mathcal{J}}_{\rm F}$ of Ref.~\cite{Becchetti:2025qlu} satisfy the non-canonical DEs in \cref{eq:DEs,eq:dM}, they were chosen such that the following terms in the derivative of ${\cal J}_{{\rm F},i}^{(w)}(\vec{x})$ vanish:
\begin{align} 
\bigl[ \dd {\rm M}^{({\rm F},0)} (\vec{x}) \cdot \vec{\cal J}_{\rm F}^{(w)} (\vec{x}) \bigr]_i & =0 \, , \label{eq:vanishing_Omega0}\\ 
\bigl[ b_{1 \beta}^{({\rm F})} \cdot \vec{\cal J}_{\rm F}^{(w-1)} (\vec{x}) \bigr]_i &=0 \, , \label{eq:vanishing_Omega1}\\ 
\bigl[ \dd {\rm M}^{({\rm F},2)} (\vec{x}) \cdot \vec{\cal J}_{\rm F}^{(w-2)} (\vec{x})\bigr]_i &=0 \, , \label{eq:vanishing_Omega2}
\end{align}
for any $i$ and $w<4$~---~i.e.\ for all terms that contribute to the poles of the two-loop amplitude~---~and, at $w=4$, for all values of $i$ except for those corresponding to MIs that are associated with either elliptic curves or nested square roots.
We verified that these conditions are numerically satisfied at several random phase-space points.
\Cref{eq:vanishing_Omega0,eq:vanishing_Omega2} imply that $\mathcal{J}_{{\rm F},i}^{(w)}$ can be written in terms of iterated integrals, whereas \cref{eq:vanishing_Omega1} means that such iterated integrals involve only logarithmic one-forms.
As a result, we can extract a set of algebraically independent functions from the set of all $\mathcal{J}_{{\rm F},i}^{(w)}$ that satisfy \cref{eq:vanishing_Omega0,eq:vanishing_Omega1,eq:vanishing_Omega2}.
We do so across all relevant one- and two-loop families ${\rm F}$ following the algorithm of Refs.~\cite{Chicherin:2020oor,Chicherin:2021dyp,Abreu:2023rco}.
We denote the resulting independent special functions by $f_i^{(w)}(\vec{x})$, where $w=1,2,3,4$ labels the transcendental weight.
Only these functions appear in the $\eps$ poles of the two-loop amplitude, thus implying that the UV and IR poles can be subtracted exactly.

We do not have an iterated-integral representation for the small subset of MI $\eps$-expansion terms $\mathcal{J}_{{\rm F},i}^{(4)}$ whose derivatives do not satisfy the conditions in \cref{eq:vanishing_Omega0,eq:vanishing_Omega1,eq:vanishing_Omega2}. 
Consequently, we lack analytic control over the relations among these functions. Instead, we search for $\mathbb{Q}$-linear relations among their numerical values using the PSLQ algorithm~\cite{pslq}. We find only the trivial relations implied by symmetries of the underlying MIs and identify a basis of $\mathbb{Q}$-linearly independent functions, denoted by $f_i^{(4^*)}(\vec{x})$. The asterisk indicates that these are not pure functions and that the superscript $4$ refers to the order in $\eps$, rather than to the transcendental weight.
Since our analysis is restricted to $\mathbb{Q}$-linear relations, we cannot exclude the existence of more complicated identities among the functions $\{f_i^{(4^*)} \}$. In this sense, this subset of functions may be overcomplete. 
Nevertheless, as shown in \cref{tab:numbers_special_functions}, their number is small. 
Moreover, the significant amplitude simplifications discussed in \cref{sec:amplitude_coefficients}, together with the cancellations reported in \cref{tab:numbers_special_functions}, confirm the usefulness of this special-function representation of the MIs.
Taken together, all MI $\eps$-expansion terms, $\mathcal{J}_{{\rm F},i}^{(w)}(\vec{x})$, can be expressed as polynomials in the special functions $\{f_i^{(w)}(\vec{x})\}$, with $w \in \{1,2,3,4,4^*\}$, and the transcendental constants ($\zeta_2$ and $\zeta_3$).
For example, we spell out a few representative terms in the expression of the $13^{\text{th}}$ MI of family ${\rm F}_1$:
\begin{align}
\begin{aligned}
 \mathcal{J}_{{\rm F}_1, 13} & = \frac{1}{48} + \frac{\eps}{24} \Bigl[f^{(1)}_1 - 2 f^{(1)}_3 + 2 f^{(1)}_4 - 2 f^{(1)}_6 \Bigr] - \frac{\eps^2}{192} \Bigl[ 28 f^{(2)}_1 - 4 \bigl(f^{(1)}_6\bigr)^2 + 8 f^{(1)}_4 f^{(1)}_6 + 11 \zeta_2 + \ldots \Bigr] \\
 & \phantom{=} \ - \frac{\eps^3}{144} \Bigl[ 36 f^{(3)}_2 - 24 f^{(3)}_{57} - 30 \bigl(f^{(1)}_1\bigr)^2 f^{(1)}_3 - 36 f^{(1)}_4 f^{(2)}_1 + 27 \zeta_2 f^{(1)}_1 + 539 \zeta_3 + \ldots \Bigr] + \eps^4 f^{(4^*)}_3 + \mathcal{O}\left(\eps^5\right) \,,
\end{aligned}
\end{align}
where we omitted the arguments for simplicity.
The complete expressions can be found in our ancillary files~\cite{zenodo}.

\begin{table}[t]
\centering
\begin{tabular}{ccccccc}
\toprule
$w$ & 1 & 2 & 3 & 4 & $4^*$ & All\\
\midrule
Bare amplitudes & 7 & 12 & 63 & 212 & 29 & 323 \\
Finite remainders & 7 & 12 & 63 & 187 & 29 & 298\\
\bottomrule
\end{tabular}
\caption{Number of special functions $f_i^{(w)}(\vec{x})$ of each order appearing in the projected bare amplitudes (truncated at order $\eps^0$) and finite remainders at two-loop order.}
\label{tab:numbers_special_functions}
\end{table}

\begin{table}[t]
\centering
\begin{tabular}{ccccccc}
\toprule
\multirow{2}{*}{Family} & Square & Rational & Algebraic & All     & Non-$\dd \log$ & \multirow{2}{*}{Functions} \\
                        & roots  & letters  & letters   & letters & one-forms      &                            \\
\midrule
${\rm A}^{\textrm{(x12)}}$ & 5 & 26 & 21 & 47 & 0 & 53\\
${\rm F}_1$ & 8 & 42 & 32 & 74 & 74 & 239\\
${\rm F}_1^{\textrm{(x12)}}$ & 6 & 35 & 26 & 61 & 74 & 148\\
${\rm F}_2$ & 10 & 36 & 37 & 73 & 64 & 178\\
${\rm F}_2^{\textrm{(x12)}}$ & 10 & 37 & 36 & 73 & 64 & 195\\
${\rm F}_2^{\textrm{(x12x34)}}$ & 7 & 31 & 27 & 58 & 42 & 124\\
${\rm F}_3$ & 7 & 29 & 28 & 57 & 0 & 127\\
\midrule
All auxiliary functions & 13 & 52 & 48 & 100 & 138 & \\
Master integrals & 13 & 65 & 61 & 126 & 220 & \\
\bottomrule
\end{tabular}
\caption{Above, we give a summary of the analytic structures appearing in the DEs for the auxiliary special functions of each family.
The last column lists the dimension of the corresponding DEs.
The last two rows report the number of distinct structures that appear in the seven DEs above, and in the DEs for the MIs of all the families listed above.}
\label{tab:analytic_structures_amplitude}
\end{table}

The set of special functions $\{ f^{(w)}_i \}$ is, by construction, closed under differentiation.
Since we have also resolved the polynomial relations among the MI $\eps$-expansion coefficients, the derivatives of each function in the set are generally given by polynomials in the functions themselves, with coefficients given by $\mathbb{Q}$-linear combinations of one-forms.
These non-linear DEs may be read off from the linear DEs for the MIs given in \cref{eq:DEs,eq:dM}.
For instance, we give here a few illustrative terms of the differential of $f_9^{(4^*)}(\vec{x})$:
\begin{equation} \label{eq:df9}
\begin{split}
\mathrm{d} f_9^{(4^*)} & = \omega^{(r_6)}_1 f_8^{(4^*)} + \omega^{(r_6)}_2 f_{10}^{(4^*)} + \frac{1}{4} \bigg( \frac{1}{2} \mathrm{d}\log W_{32}^{(r_6)} + \frac{7}{3} \omega^{(r_6)}_3 \bigg) f_{49}^{(3)} \\
& \phantom{=} \ + \mathrm{d}\log W_{28}^{(r_6)} \Big[ 4 f_2^{(1)} f_{12}^{(2)} -4 \bigl( f_2^{(1)} \bigr)^3 -\frac{3}{4} \zeta_2 f_5^{(1)} \Bigr] + \ldots \, ,
\end{split}
\end{equation}
where we omitted the argument $\vec{x}$ for simplicity, and the superscript $r_6$ indicates that the one-forms $\omega^{(r_6)}_i$ and $\mathrm{d}\log W_{i}^{(r_6)}$ are odd with respect to flipping the sign of the square root $r_6$ defined in Ref.~\cite{Becchetti:2025qlu} and in our ancillary files~\cite{zenodo}.
In order to employ the publicly available tools for the solution of linear DEs via series expansions, as discussed in \cref{sec:numerical_eval}, we enlarge the set of special functions to include also the special-function polynomials appearing in the derivatives (see e.g.\ the term in the square brackets on the right-hand side of \cref{eq:df9}).
The DEs for the special functions offer a number of advantages over those for the MIs: they are $\eps$-independent and sparser (see Ref.~\cite{Badger:2024dxo} for a visual representation of the typical structure), and some analytic structures appearing in the DEs for the MIs drop out of those for the special functions, as illustrated in \cref{tab:analytic_structures_amplitude}.
On the other hand, the DEs for the complete set of special functions have a larger dimension than those for each set of MIs separately.
To mitigate this growth, we evaluate each integral family independently.
We therefore construct, for each family, an \textit{auxiliary} basis of special functions, and solve the corresponding DEs.
The \textit{global} set of functions $\{f^{(w)}_i(\vec{x})\}$, enumerated in \cref{tab:numbers_special_functions}, is then obtained by expressing it in terms of these (redundant) auxiliary functions.
The relation between the global and the auxiliary sets is not one-to-one, as the same global function can appear in the auxiliary basis for different families. 
Therefore, we include in the DEs for the auxiliary functions of each family only the minimal set of functions needed to evaluate the global basis.
In doing so, we preferably choose for each global $f_i^{(w)}$ an auxiliary representative drawn from the simplest possible family.
For example, the MIs of both ${\rm F}_2$ and ${\rm F}_3$ contain the functions $ f^{(4^*)}_{11}, \dots, f^{(4^*)}_{18}$; we evaluate these through the auxiliary functions of ${\rm F}_2$, and omit the corresponding functions from the auxiliary set of ${\rm F}_3$.
The number of needed auxiliary special functions for each family is given in \cref{tab:analytic_structures_amplitude}.
Lastly, since the computation of the boundary values using \textsc{AMFlow} is expensive,\footnote{
\label{fn:amflow}For reference, evaluating the MIs of ${\rm F}_2$, the most expensive family, required 30--40 hours of wall-clock time per point on 21 parallel threads, using \textsc{AMFlow} with the IBP reductions performed by \textsc{FiniteFlow} and \textsc{LiteRed}~\cite{Lee:2013mka}.} we only compute them for the two-loop families ${\rm F}_1$, ${\rm F}_2$ and ${\rm F}_3$, and then evaluate the permuted families by integrating the unpermuted DEs to the permuted target point.
We stress that no analytic continuation across branch-cut singularities is required, because the needed permutations map the physical region defined in \cref{sec:conventions} onto itself.
The solution of the DEs is discussed in the next subsection.

\section{Numerical evaluation and results}
\label{sec:results}
\label{sec:numerical_eval}

In this section, we discuss the numerical evaluation of the special functions introduced in the previous section.
We combine these with the numerical values of the rational coefficients, obtained as described in \cref{sec:amplitude_coefficients}, to evaluate the finite projected amplitudes in~\cref{eq:preFFfin} and, through \cref{eq:interferences_B}, the hard functions that contribute to the cross section.
We then present the evaluation of the hard functions on the phase-space grid employed to compute the NNLO QCD cross section in Ref.~\cite{Becchetti:2026awn}.
Finally, we discuss the checks we performed and provide benchmark values.

We evaluate the special functions by numerically solving the systems of DEs described in~\cref{sec:special_functions}. To this end, we use the DE solver of \textsc{AMFlow}~\cite{Liu:2022chg}, which provides a \textsc{Mathematica} implementation of the method of generalised power series~\cite{Pozzorini:2005ff,Moriello:2019yhu}.
Given known values at a starting point, the solver transports the solution to the target point along a straight line in the space of invariants $\vec{x}$ by constructing and matching local generalised power-series solutions on consecutive segments.
A complication arises because the physical region defined in \cref{sec:conventions} is not convex in the space of invariants $\vec{x}$. Consequently, the straight-line integration path connecting a starting point to a target point may leave the physical region, thereby crossing branch cuts of the special functions and requiring analytic continuation. 
To avoid this complication, we choose a set of starting points such that, for each target point, there exists at least one starting point for which the straight-line path to the target remains entirely within the physical region.
For the application discussed below, we employ a set of $21$ starting points, where the special functions are precomputed with at least $35$ significant digits using \textsc{AMFlow}. 
We selected the starting points from a pool of candidates consisting of the grid points, supplemented by a few phase-space points at which the MIs had previously been evaluated for validation purposes. 
From this pool, we determined a small set of starting points as follows. 
For each point in the pool, we identified the subset of grid points connected to it by a straight path lying entirely within the physical region. 
We then selected as the first starting point the candidate connected to the largest number of grid points. 
Finally, we removed from the pool all points connected to the starting points selected so far, and repeated the selection on the remaining points until the entire grid was covered.

Even when the path lies entirely within the physical region, the DEs may exhibit spurious poles, i.e.\ singularities of the DEs at which the solution remains finite.
Since these singularities are unphysical, they may be bypassed without ambiguity.
In practice, we perform each integration twice, varying both the truncation order of the series expansions and the prescription used to bypass the spurious singularities, namely, whether the contour is deformed into the upper or lower half of the complex plane.
We then require the two results to agree to better than $10^{-16}$ and, if necessary, increase the expansion orders until this condition is met.
This provides a stringent check that the encountered singularities are indeed spurious and that the numerical integration has converged.
Although this procedure requires two integrations, the computational cost is less than doubled because the parametrisation of the DEs along the path, which constitutes a significant fraction of the runtime, is performed only once.
In terms of total CPU time, our routines evaluate the special functions roughly two orders of magnitude faster than the \textsc{AMFlow} setup we used to compute the master integrals at the starting points (cfr.\ \cref{fn:amflow}).
Nonetheless, the evaluation of the special functions dominates the runtime of the evaluation of the finite projected amplitudes.

We evaluated the hard function up to two-loop order on a grid of 224`640 phase-space points using five machines with 100 threads each.
This grid was used to interpolate the two-loop hard function with quadratic splines, and compute the NNLO QCD cross section in the LCA in Ref.~\cite{Becchetti:2026awn}.
Parametrising the momenta as in Ref.~\cite{Agarwal:2024jyq}, the grid is defined in terms of two energy fractions ($\beta^2$, $\operatorname{frac}_{s_{t\bar t}}$) and three angular variables ($\cos \theta_W$, $\cos \theta_t$, $\phi_t$):
\begin{align}
 \vec{\alpha} = \bigl( \beta^2 \,, \operatorname{frac}_{s_{t\bar t}} \,, \cos \theta_W \,, \cos \theta_t \,, \phi_t \bigr) \,.
\end{align}
The energy fractions are defined as
\begin{equation} \label{eq:physvar}
\beta^2 = 1 - \frac{(2m_t + m_W)^2}{s_{12}} \,, \qquad \quad \operatorname{frac}_{s_{t\bar{t}}} = \frac{s_{34} - 4 m_t^2}{(\sqrt{s_{12}}-m_W)^2 - 4 m_t^2} \,,
\end{equation}
$\theta_W$ is the polar angle of the $W$ boson in the partonic centre-of-mass frame, while $\theta_t$ and $\phi_t$ are, respectively, the polar and azimuthal angles of the top quark in the $t \bar t$ rest frame.
We refer to Appendix~A.4 of Ref.~\cite{Agarwal:2024jyq} for a detailed discussion of this parametrisation.\footnote{The momenta $q_i$ parametrised in Ref.~\cite{Agarwal:2024jyq} are related to the momenta $p_i$ in \cref{eq:ScatteringProcess} by $q_1=p_1$, $q_2=p_2$, $q_3=-p_3$, $q_4=-p_4$, and $q_5=-p_5$.}
The masses are fixed to $m_t = 173.2 \, \mathrm{GeV}$ and $m_W = 80.385 \, \mathrm{GeV}$.
The physical region then corresponds to the following variable ranges:
\begin{equation} \label{eq:varbound}
\beta^2 \in \left[0,1\right] \,, \qquad \operatorname{frac}_{s_{t\bar t}} \in \left[0,1\right] \,, \qquad \cos \theta_W \in \left[-1,1\right] \,, \qquad \cos \theta_t \in \left[-1,1\right] \,, \qquad \phi_t \in \left[0, 2 \pi \right] \,.
\end{equation}
The grid is the tensor product of five non-uniform one-dimensional grids containing $20$, $18$, $13$, $8$ and $6$ points in the variables $\vec{\alpha}$, for a total of 224'640 phase-space points.
The grid excludes the endpoints of the ranges in \cref{eq:varbound} except for $\phi_t$, for which only the endpoint $\phi_t = 2 \pi$ is omitted, as it is equivalent to $\phi_t=0$.
Because the scalar invariants $\vec{x}$ depend on $\phi_t$ only through $\cos\phi_t$, grid points related by $\phi_t \to 2\pi-\phi_t$ coincide in the $\vec{x}$ space.
Removing these duplicates reduces the number of distinct phase-space points at which the finite projected amplitudes must be evaluated to 149'760.
The hard functions, on the other hand, also have a parity degree of freedom, which is entirely captured by the pseudo-scalar invariant $\mathrm{tr}_5$ in \cref{eq:mijtr5,eq:interferences_B}.
Because $\mathrm{tr}_5$ changes sign under $\phi_t\to2\pi-\phi_t$, care must be taken to assign it the correct sign, according to \cref{eq:tr5}, when reconstructing the full grid from the subset at which the projected amplitudes have been computed.

\begin{figure}[t]
    \centering
    \includegraphics[width=.6\textwidth]{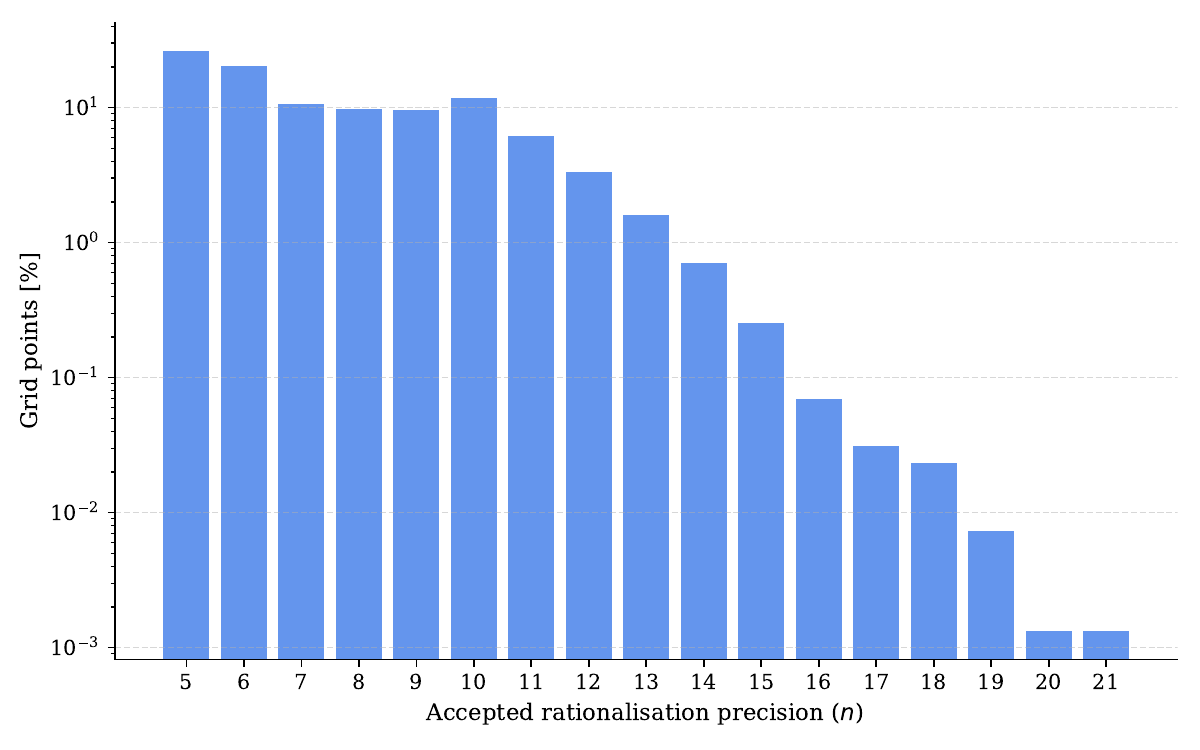}
    \caption{Distribution of the grid points as a function of the accepted rationalisation precision $n$.}
    \label{fig:histo_ratprec}
\end{figure}

A subtlety arises in the relation between the grid variables $\vec{\alpha}$ and the invariants $\vec{x}$.
The finite-field reduction procedure described in \cref{sec:amplitude_coefficients} and the tools we use to evaluate the special functions require \textit{rational} values of $\vec{x}$.
Since the transformation between $\vec{\alpha}$ and $\vec{x}$ is irrational, the rationalisation must be performed directly in the $\vec{x}$ space.
The choice of rationalisation precision involves a trade-off.
Increasing the precision requires more prime numbers for the rational reconstruction and slows down the solution of the DEs.
Decreasing it, however, moves the rationalised point farther away from the corresponding exact point in the grid variables, thereby introducing an additional source of interpolation error.
We rationalise each grid point as follows.
We start from the definition of the point in the grid variables, $\vec{\alpha}$.
We denote by $\Phi$ the map from grid variables to invariants, by $\Phi(\vec{\alpha})$ the exact, irrational image of $\vec{\alpha}$ in the invariant space, and by $\vec{x}$ the rationalisation of $\Phi(\vec{\alpha})$.
The masses~---~$x_6$ and $x_7$~---~are always rationalised as
\begin{equation}
\label{eq:rationalised_masses}
    m_t^2=\frac{749956}{25}\,\mathrm{GeV}^2\,,
    \qquad \quad
    m_W^2=\frac{258469929}{40000}\,\mathrm{GeV}^2\,.
\end{equation}
For the remaining five kinematic invariants, we iterate over the rationalisation precision $n$, starting from $n=5$.
We define $x_i$, i.e.\ the rationalisation of $\Phi(\vec{\alpha})_i$, as the rational number with the smallest denominator such that
\begin{align}
\bigl| x_i - \Phi(\vec{\alpha})_i \bigr| <  10^{-n} \times\, \mathrm{min}_{j=1,\ldots,5} \bigl|\Phi(\vec{\alpha})_j \bigr| \qquad \forall \, i=1,\ldots,5 \,.
\end{align}
We accept the rationalisation $\vec{x}$ if it belongs to the physical region (cfr.\ \cref{sec:conventions}), and if the corresponding point in the grid variables differs from the original one by less than five significant digits, i.e.\ if
\begin{align}
\mathrm{max}_{i=1,\ldots,5} \left| \frac{\bigr[\Phi^{-1}(\vec{x})\bigr]_i}{\alpha_i} - 1 \right| < 10^{-5} \,.
\end{align}
If these conditions are not met, we increase the rationalisation precision $n$ by $1$ until they are.
The distribution of the grid points with respect to their accepted rationalisation precision $n$ is shown in \cref{fig:histo_ratprec}.
For most points $n<10$ is sufficient, while a small fraction lies in the tail at larger rationalisation precision, up to $n=21$.
Another consequence of rationalising the invariants concerns the value assigned to $\mathrm{tr}_5$. 
Although the pseudo scalar defined in \cref{eq:tr5} can be computed from the grid variables $\vec{\alpha}$ through the exact momenta $\{p_i\}$, this would generally yield a value of $\mathrm{tr}_5^2$ that is inconsistent with its definition in terms of the rationalised invariants $\vec{x}$ through \cref{eq:tr5_to_invariants}.
Therefore, we evaluate $\mathrm{tr}_5$ directly from the rationalised invariants up to its sign, which is determined from the grid variables.
In the exceptional case in which $\mathrm{tr}_5$ vanishes identically when expressed in terms of the grid variables, but not in terms of the rationalised invariants, we retain the exact value $\mathrm{tr}_5=0$.
In Ref.~\cite{Becchetti:2026awn}, the impact of the rationalisation procedure outlined above was found to be negligible at the level of the inclusive NLO QCD cross section by comparison with the implementation of the one-loop amplitude of Ref.~\cite{Becchetti:2025osw}, which employs a different and less aggressive rationalisation of the invariants.
Therefore, we are confident that the chosen rationalisation precision remains a negligible source of error also at the two-loop order.

A crucial aspect of our computation is the assessment of the numerical accuracy of the two-loop hard function.
We recall that the rational coefficients of the finite projected amplitudes are exact rational numbers.
The numerical uncertainty therefore arises from the evaluation of the special functions and from possible cancellations when the finite remainders are assembled and contracted with the tree-level amplitude.
In principle, this uncertainty could be estimated by computing two independent sets of special functions at each phase-space point, obtained by integrating the DEs from different starting points, and comparing the corresponding values of the finite remainder.
Since the evaluation of the MIs at the starting points with \textsc{AMFlow} is computationally expensive, we minimised the number of starting points required to cover the phase-space grid, thereby foregoing such a direct estimate of the numerical uncertainty.
In other words, a grid point is generally reachable from a single starting point.
To overcome this limitation, we devised the following validation strategy.

We first computed the special functions using our standard settings in \textsc{AMFlow}'s DE solver, hereafter referred to as the ``standard-precision'' set. The associated numerical uncertainty was then propagated to the finite remainder by means of the \textit{replica method} (see e.g.\ Ref.~\cite{Costantini:2024wby} and references therein). More precisely, at each phase-space point, we generated $200$ replicas of the standard-precision set of special functions by independently sampling their real and imaginary parts from Gaussian distributions centred on their central values. For each special function, the width of the distribution was taken to be the maximum between the estimated uncertainty from the truncation of the local power-series expansion and the numerical precision of the \textsc{AMFlow} starting values. The resulting ensemble of finite remainders provides, for each phase-space point, an estimate of the uncertainty induced by the numerical evaluation of the special functions.

The replica analysis identified 141 grid points with relative standard deviation exceeding a chosen threshold of $10^{-6}$. 
Only for these points, the special functions were recomputed using more stringent numerical settings, including a higher working precision and deeper series expansions (the ``high-precision'' set), and the corresponding finite remainders were re-evaluated. Comparison between the standard- and high-precision results confirmed insufficient numerical stability for only 27 grid points. 
For this small subset, we could recompute the special functions by integrating the DEs from a different \textsc{AMFlow} starting point.
The comparison of these independent evaluations provided a robust estimate of the residual numerical uncertainty.
Overall, this validation strategy ensured a numerical precision of at least five significant digits for the two-loop finite remainder over the entire phase-space grid, with substantially higher accuracy for the vast majority of grid points.

We have validated our results through the following checks:
\begin{itemize}
	\item The analytic cancellation of all poles required to construct the finite remainder demonstrates consistency with the universal structure of UV and IR singularities (see \cref{sec:amplitude}).
	\item We cross-checked the one-loop hard function (up to $\mathcal{O}(\eps^2)$) against Ref.~\cite{Becchetti:2025osw}.
	To compare against the leading-colour results derived in this work, we removed the subleading-colour terms from the full-colour amplitudes of Ref.~\cite{Becchetti:2025osw} and extended their implementation to allow for a renormalisation scale $\mu$ different from $m_t$.
	We checked the latter change against \OpenLoops~\cite{Buccioni:2019sur} (up to $\mathcal{O}(\eps^0)$).
	\item We cross-checked the one- and two-loop hard functions against an independent calculation performed in CDR, as discussed in \cref{sec:cdr}.
\end{itemize}

In our ancillary files~\cite{zenodo}, we provide the values of the leading-colour two-loop hard function (cfr.\ \cref{eq: hard-function}) evaluated at the 224'640 phase-space points of the grid discussed above.
To facilitate validation, \cref{tab:benchmarks} lists benchmark values, decomposed into the relevant $(N_c,n_l)$ terms, at three representative points of the~grid:
\begin{subequations}
\label{eq:BPs}
\begin{align}
	\vec{x}_{\rm BP1} & = \biggl( -\frac{22232616352}{506225} \,, \frac{18820536029}{156847} \,, -\frac{76575190888}{1742729}\,, -\frac{16183193820}{581627} \,, -\frac{15773715997}{566041} \biggr) \,, \\
	\vec{x}_{\rm BP2} & = \bigg(-\frac{262259}{6} \,, \frac{1080869}{9} \,, -\frac{820621}{19} \,, -\frac{141893}{5} \,, -\frac{191736}{7} \biggr) \,, \\
	\vec{x}_{\rm BP3} & = \biggl(-\frac{4178919078344}{95114953} \,, \frac{12301151022663}{102515596} \,, - \frac{6150018491354}{139952799} \,, -\frac{12608908588612}{452949967} \,, -\frac{4327639738910}{155371909} \biggr) \,,
\end{align}
\end{subequations}
in units of $\mathrm{GeV}^2$, where we omit the last two elements of $\vec{x}$, namely the masses, as they are fixed to the values in \cref{eq:rationalised_masses}.
The pseudo-scalar invariant $\mathrm{tr}_5$ is computed from the invariants through \cref{eq:tr5_to_invariants}, with its sign fixed as
\begin{align} \label{eq:BPtr5}
 \operatorname{Im} \mathrm{tr}_5(\vec{x}_{\rm BP1}) > 0 \,, \qquad \operatorname{Im} \mathrm{tr}_5(\vec{x}_{\rm BP2}) < 0 \,, \qquad \operatorname{Im} \mathrm{tr}_5(\vec{x}_{\rm BP3}) > 0 \,.
\end{align}
The renormalisation scale $\mu$ is set to the invariant mass of the event, i.e.\ $\mu = \sqrt{s_{12}}$.

\begin{table}[t]
\centering
\begin{tabular}{l c c c}
\toprule
\multicolumn{1}{c}{$[\mathrm{GeV}^{-2}]$} &
\multicolumn{1}{c}{BP1} &
\multicolumn{1}{c}{BP2} &
\multicolumn{1}{c}{BP3} \\
\midrule
$\Hard^{(0)}$ & 0.0001237832195483 & 0.0001235499345212 & 0.0001237832002864 \\
\midrule
$\Hard^{(1)}_{\{1,0\}}$ & 0.0018829581333315 & 0.0018854539381310 & 0.0018828565414323  \vspace{0.07cm}\\
$\Hard^{(1)}_{\{0,1\}}$ & -0.0003439587672146 & -0.0003433331715764  & -0.0003439586994571 \vspace{0.07cm}\\
$\Hard^{(1)}$ & 0.0039290805639217 & 0.0039396959565109 & 0.0039287761270113 \\
\midrule
$\Hard^{(2)}_{\{2,0\}}$ & 0.03329229368 & 0.0332529672386551 & 0.0332904198 \vspace{0.07cm}\\
$\Hard^{(2)}_{\{1,1\}}$ &-0.00758024593  &-0.0075740277064554  & -0.0075801539 \vspace{0.07cm}\\
$\Hard^{(2)}_{\{0,2\}}$ &-0.00060806556  & -0.0006068566788678 & -0.0006080655 \vspace{0.07cm}\\
$\Hard^{(2)}$ & 0.17072531514(6) & 0.1704948725793703(8) & 0.1707098315(2) \\
\bottomrule
\end{tabular}
\caption{Numerical results for the Born, one-loop and two-loop hard functions for the benchmark phase-space points in \cref{eq:BPs,eq:BPtr5}.
We strip off the overall coupling-dependent factor of $g_w^2 (4 \pi \as)^2 (\as/(4 \pi) )^{\ell}$ from $\Hard^{(\ell)}$ in \cref{eq: hard-function}, and give the separate terms $\Hard^{(\ell)}_{\{a,b\}}$ of the $(N_c,n_l)$ decomposition (cfr.\ \cref{eq: colour-decomposition}).
The error in brackets assigned to the full two-loop results is estimated as discussed in the main text.
}
\label{tab:benchmarks}
\end{table}

\section{Calculation in the conventional dimensional regularisation scheme}
\label{sec:cdr}

In addition to the computation in the 't Hooft-Veltman scheme discussed in~\cref{sec:thv}, we have performed an independent calculation of the two-loop amplitudes in the conventional dimensional regularisation scheme.
This serves as a cross-check of the entire computational framework.
The strategy adopted here differs from that described in~\cref{sec:thv} in several key aspects, which we detail below.

%%%%%%%%%%%%%%%%%%%%%%%
\subsection{Amplitude construction}
\label{sec:cdr_amplitude}

The Feynman diagrams contributing to the partonic process in~\cref{eq:ScatteringProcess} are generated with \textsc{FeynArts}~\cite{Hahn:2000kx}.
The interference between the two-loop colour-dressed amplitude and the tree-level counterpart is computed using a private \textsc{Mathematica} package, which performs the contraction of both colour and Lorentz indices and the Dirac algebra.
This directly provides the unrenormalised polarisation-summed interferences that contribute to the definition of the hard function in~\cref{eq: hard-function}.
The approach thus bypasses the decomposition into tensor structures and the use of physical projectors described in~\cref{sec:integrand_generation}.

%%%%%%%%%%%%%%%%%%%%%%%
\subsection{Integral reduction with block-triangular systems}
\label{sec:cdr_reduction}

The reduction of the scalar loop integrals to MIs is performed with a substantially different strategy from the one described in~\cref{sec:amplitude_coefficients}.

To reduce the algebraic complexity of the integral reduction, we fix the top-quark and $W$-boson masses to numerical values normalised to the top-quark mass, which amounts to setting $m_t=1$.
In addition, throughout the entire calculation, the dimensional regulator $\eps$ is assigned the numerical values $\pm \bar{\eps}$, with $\bar{\eps}=10^{-3}$~\cite{Bi:2023bnq,Bi:2025oga,Li:2025bsq}.
As a consequence, only the five independent Mandelstam invariants $s_{ij}$ appearing in $\vec{x}$ of~\cref{eq:invariants} are kept symbolic in the generation of the reduction identities.
This drastically reduces the algebraic complexity compared to the fully symbolic reduction, at the cost of performing numerical evaluations at two values of $\eps$.

We generate the full IBP system using the package \textsc{Blade}~\cite{Guan:2024byi}, where the MIs are chosen according to the Laporta algorithm~\cite{Laporta:2000dsw}.
Rather than using this full IBP system directly to generate samples for reconstruction, we first solve this system over prime fields on a small set of sample phase-space points.
This small set of reductions is then used as input to search for a \textit{block-triangular form}~\cite{Guan:2019bcx}, namely a much smaller system of linear relations among the required target integrals and the MIs, which does not involve the large number of auxiliary integrals present in the full IBP system and is characterised by a block-by-block structure.
For the most complicated integral family, which corresponds to ${\rm F}_1$ in~\cref{tab:Intfam1}, the search and reconstruction of this block-triangular form takes approximately one week on 20 CPU cores.

Once the block-triangular form is available, the conventional strategy would be to use it in place of the full IBP system to generate a large number of finite-field samples for the exact rational reconstruction described in~\cref{sec:amplitude_coefficients}.
The novel aspect of our approach is that, instead, we evaluate the reduction coefficients using floating-point arithmetic in \textsc{Mathematica} at each phase-space point.
The block-by-block structure of the block-triangular system enables a recursive solution from the bottom to the top sector~\cite{Guan:2019bcx}.
At each step, the coefficient matrix on the left-hand side consists of rational numbers, because the kinematic variables and $\eps$ have been fixed to the given phase-space point.
The right-hand side is a linear combination of integrals belonging to lower sectors, whose numerical values have already been determined in previous iterations.
The target integrals in the current block are then obtained by inverting the coefficient matrix and multiplying it by the right-hand-side vector.
The boundary values for the lowest block are provided by the MIs.

The use of floating-point arithmetic in this context rests on two observations.
First, the block-triangular form is remarkably compact compared to the full IBP system, precisely because it does not contain the large number of auxiliary integrals present in the IBP identities.
Taking the most complicated family as an example, the block-triangular system consists of 3844 linear equations, which is fewer by an order of magnitude than the number of equations in the full IBP system.
Second, the size of each individual block within the block-triangular system is of order $\mathcal{O}(100)$, so that the accumulation of numerical errors from the recursive solution remains negligible.
The dominant source of the overall numerical uncertainty does not originate from this step, but rather from the chosen numerical values for $\eps$ (see \cref{sec:cdr_evaluation}).

%%%%%%%%%%%%%%%%%%%%%%%
\subsection{Numerical evaluation and cross-check}
\label{sec:cdr_evaluation}

The MIs are evaluated by numerically solving the systems of DEs taken from Ref.~\cite{Becchetti:2025qlu} using a one-dimensional parametrisation along straight-line paths.
In contrast to the approach of~\cref{sec:special_functions}, where the MIs are expanded in $\eps$ and expressed in terms of special functions, here we directly integrate the $\eps$-dependent DEs using the numerical values $\pm \bar{\eps}$.
We employ the DE solver of \textsc{AMFlow}~\cite{Liu:2017jxz,Liu:2021wks,Liu:2022chg} with the same setup described in~\cref{sec:numerical_eval}.
The boundary values are computed with \textsc{AMFlow} to at least 35-digit precision.

Note that, since the initial Laporta basis produced by \textsc{Blade} and the basis adopted in Ref.~\cite{Becchetti:2025qlu} are different, the IBP reduction coefficients relating the two integral bases must also be determined.
This can be achieved using the same block-triangular and floating-point strategy described above.

After performing the UV renormalisation and IR subtraction according to the procedure of~\cref{sec:polestructure}, the numerical squared matrix element at a given value of $\eps$ can be expanded as the sum of the hard function and higher-order terms in $\eps$.
Evaluating the hard function at $\eps = \pm \bar{\eps}$ and taking their average,
\begin{equation}
\frac{\Hard^{(2)}(+\bar{\eps}) + \Hard^{(2)}(-\bar{\eps})}{2} = \Hard^{(2)}(0) + \mathcal{O}(\bar{\eps}^2) \,,
\end{equation}
we obtain the physical hard function, $\Hard^{(2)}(0)$, with an uncertainty of order $\bar{\eps}^2 \sim 10^{-6}$.

We cross-checked the two-loop hard function obtained with this CDR-based approach against the result of the tHV computation discussed in~\cref{sec:thv} at three representative phase-space points (see \cref{tab:benchmarks}).
The two determinations agree within five significant digits, which is consistent with the expected numerical uncertainties of both calculations.
This independent computation thus provides a robust validation of the complete computational framework.

\section{Conclusions}
\label{sec:conclusions}

In this work, we have presented a numerical computation of the two-loop QCD amplitudes for $\ttW$ production in the generalised leading-colour approximation, retaining the full dependence on the top-quark and $W$-boson masses.
The combination of $2\to 3$ kinematics and multiple massive internal and external particles makes $\ttW$ production one of the most challenging applications of current multi-loop amplitude techniques. 

Rather than pursuing a fully analytic calculation, we adopted a hybrid strategy that combines numerical evaluation with strong analytic and algebraic control throughout the calculation.
We numerically evaluate the two-loop amplitude in such a way that many cancellations occur exactly rather than in floating-point arithmetic.
To this end, we expressed the master integrals in terms of a set of special functions; most of these are algebraically independent, while for the few functions encoding the most intricate analytic structures~---~elliptic curves and nested square roots~---~we verified only their $\mathbb{Q}$-linear independence.
This representation allows the two-loop finite remainder to be extracted with UV and IR poles cancelling exactly, while also exposing algebraic cancellations that significantly simplify the calculation.
The finite remainder is ultimately expressed in terms of these (mostly) algebraically independent special functions and linearly independent rational coefficients.
We evaluate the special functions by numerically solving systems of differential equations with series expansions.
Rather than reconstructing the analytic expression of the rational coefficients, we evaluate them by reconstructing their exact rational values from finite-field evaluations at each target phase-space point.
Using this framework, we evaluated the two-loop hard function on a grid of 224'640 points, provided in our ancillary files~\cite{zenodo}.

We validated our results by performing a second independent calculation of the two-loop amplitude, based on a different regularisation scheme (CDR) and a substantially different computational strategy.
The CDR implementation computes the interference with the Born amplitude directly, rather than relying on tensor projectors, employs an independent strategy for the reduction of the loop integrals without introducing a special-function decomposition, and evaluates the hard function at numerical values of the dimensional regulator $\eps$.

The results presented here have already been employed to obtain the double virtual contribution to the NNLO QCD $\ttW$ cross section~\cite{Becchetti:2026awn},
demonstrating that the combination of the strategy developed in this work with interpolation techniques is suitable for phenomenological applications.
More broadly, we believe that our hybrid approach, together with ongoing progress in interpolation methods aimed at reducing the required number of evaluations~\cite{Breso-Pla:2024pda}, will enable NNLO predictions for a range of processes for which fully analytic two-loop amplitude calculations remain out of reach.

%============================================
\vspace{0.5cm}
\noindent {\bf Acknowledgments}\\
\noindent
We are grateful to Massimiliano Grazzini, Stefan Kallweit and Lorenzo Tancredi for valuable discussions and for helpful comments on the manuscript.
We thank Alessandro Porcari for useful comments on the draft.
This work has been supported by the European Research Council under the European Union’s Horizon Europe research and innovation programme through grants No.~101040760 (ERC Starting Grant \emph{FFHiggsTop}: M.B., D.C., V.C., T.P., and M.P.), 
No.~101167287 (ERC Synergy Grant \emph{MaScAmp}: V.C.), No.~949279 (ERC Starting Grant \emph{HighPHun}: M.D. and S.D.), No.~101044599 (ERC Consolidator Grant \emph{JANUS}: M.D.), and No.~101118787 (ERC Synergy Grant \emph{UNIVERSE PLUS}: S.D.).
Views and opinions expressed are however those of the author(s) only and do not necessarily reflect those of the European Union or the European Research Council Executive Agency. Neither the European Union nor the granting authority can be held responsible for them.
The work of X.C.~is supported by the Swiss National Science Foundation (SNSF) under contract 200020$\_$219367 and partly by the UZH Postdoc Grant, grant No.~[FK-25-104].
The work of C.S.~is partly supported by the Excellence Cluster ORIGINS, funded by the Deutsche Forschungsgemeinschaft (DFG, German Research Foundation) under Germany’s Excellence Strategy~---~EXC-2094-390783311.
S.Z.~was supported by the Swiss National Science Foundation (SNSF) under the Ambizione grant No.~215960.\\

\appendix
\section{Renormalisation factors and anomalous dimensions}
\label{sec:appendixA}

In this appendix we list the renormalisation constants entering the subtraction of UV singularities according to \cref{eq:UVren_amplitudes}, as well as the anomalous dimensions appearing in the subtraction of the IR poles in \cref{eq: anomalous_dimension}.

We begin with the one- and two-loop strong coupling renormalisation constants in \cref{eq: coupling_ren_constant}, given by
\begingroup
\allowdisplaybreaks
\begin{align}
	\delta Z_{\as}^{(1)} &= -\frac{\beta_0}{\eps} \,,\\
	\delta Z_{\as}^{(2)} &= \frac{\beta_0^2}{\eps^2} -\frac{\beta_1}{2 \eps} \,,
\end{align}
\endgroup
where the beta-function coefficients are
\begingroup
\allowdisplaybreaks
\begin{align}
	\beta_0 &= \frac{11}{3} C_A - \frac{4}{3} T_F n_l \,,\\
	\beta_1 &= \frac{34}{3} C_A^2 - \frac{20}{3} C_A T_F n_l - 4 \, C_F T_F n_l \,,
\end{align}
\endgroup
with $T_F = 1/2$, and
\begin{equation}
	C_A = N_c\,, \qquad  C_F = \frac{N_c}{2} \,,
\end{equation}
in the leading-colour approximation.
The expansions of the top-quark mass and wavefunction renormalisation constants are
\begingroup
\allowdisplaybreaks
\begin{align}
	Z_t &= 1 + \frac{\as(\mu)}{4 \pi} \delta Z_t^{(1)} +  \left(\frac{\as(\mu)}{4 \pi}\right)^2 \delta Z_t^{(2)} + \mathcal{O}(\as^3)  \,, \\
	Z_m &= 1 + \frac{\as(\mu)}{4 \pi} \delta Z_m^{(1)} + \mathcal{O}(\as^2)  \,,
\end{align}
\endgroup
where the perturbative coefficients are given in the on-shell renormalisation scheme by~\cite{Broadhurst:1991fy,Barnreuther:2013qvf}
\begingroup
\allowdisplaybreaks
\begin{align}
	& \delta Z_t^{(1)} = -C_F \left(\frac{3}{\eps}+\frac{4}{1-2 \eps}\right) \left(\frac{\mu^2}{m_t^2}\right)^{\!\eps} e^{\eps \gamma_E}  \Gamma(1+\eps) \,,\\
	& \begin{aligned}
	\delta Z_t^{(2)} &= \delta Z_t^{(1)} \delta Z_{\as}^{(1)} +  \left(\frac{\mu^2}{m_t^2}\right)^{\!2 \eps} \!\!e^{2 \eps \gamma_E} \Gamma(1+\eps)^2
\biggl\{
C_F T_F n_l \left(\frac{2}{\eps^2}+\frac{9}{\eps}+\frac{59}{2}+8 \, \zeta_2 \right) \\
&\phantom{=} \ + C_F^2 \left(\frac{9}{2 \eps^2}+\frac{51}{4 \eps}+ \frac{433}{8}-78 \, \zeta_2+96 \, \zeta_2 \log(2) -24 \, \zeta_3\right) \\
&\phantom{=} \ + C_A C_F \left(-\frac{11}{2 \eps^2}-\frac{101}{4 \eps}-\frac{803}{8}+30 \, \zeta_2-48 \, \zeta_2 \log(2) +12 \, \zeta_3 \right)
\biggr\} \,,
\end{aligned}
\end{align}
\endgroup
and
\begingroup
\allowdisplaybreaks
\begin{align}
\begin{aligned}
	\delta Z_m^{(1)} & = -C_F \biggl\{
	\frac{3}{\eps} + 4 + 3 \, l_{\mu} + \eps\left( 8 + \frac{3}{2}\zeta_2 + \frac{3}{2} l_{\mu}^2 + 4 \, l_{\mu}  \right) \\
	& \phantom{=} \ + \eps^2 \biggl[ 16 + 2 \, \zeta_2 - \zeta_3 + \frac{l_{\mu}^3}{2} + 2 \, l_{\mu}^2 + l_{\mu}\left(8 + \frac{3}{2}\zeta_2\right) \biggr] + \mathcal{O}(\eps^3)
	\biggr\}\,,
\end{aligned}
\end{align}
\endgroup
with $l_{\mu} = \log(\mu^2/m_t^2)$\,.

Finally, we provide the LC expression of the cusp and collinear anomalous dimensions~\cite{Becher:2009qa,Becher:2009kw} in \cref{eq: anomalous_dimension}:
\begingroup
\allowdisplaybreaks
\begin{align}
\gamma_{\mathrm{cusp}}^{(1)} &= 4 \,, \\
\gamma_{\mathrm{cusp}}^{(2)} &= \left(\frac{268}{9} -8 \, \zeta_2 \right) C_A -\frac{80}{9}T_F n_l \,, \\
\gamma_q^{(1)} &= -3 \, C_F,
\\
\gamma_q^{(2)} &= C_F^2 \left(-\frac{3}{2}+12 \, \zeta_2 -24 \, \zeta_3\right) + C_F C_A \left(-\frac{961}{54} - 11 \, \zeta_2 +26 \, \zeta_3 \right) +C_F T_F n_l \left( \frac{130}{27} + 4\zeta_2 \right) \,, \\
\gamma_{Q}^{(1)} &= -2\, C_F \,, \\
\gamma_{Q}^{(2)} &= C_F C_A \left( 4 \, \zeta_2 -\frac{98}{9} -4 \, \zeta_3 \right) +\frac{40}{9}C_F T_F n_l \,.
\end{align}
\endgroup

\section{Tensor basis}
\label{app:appendixB}

In this appendix, we give the basis of Lorentz tensors $T_i$ used in the form-factor decomposition of the partial amplitudes in \cref{eq:tensordecomposition}.
The tensors $T_i$ take the general form
\begin{equation}
\label{eq:tensorbasis}
\left( \overline{v}_1 \Gamma_0 u_2 \right) \,  \left( \overline{V}_3\Gamma_m U_4 \right) \, \kappa \,,
\end{equation}
where $\Gamma_0$, $\Gamma_m$ and $\kappa$ are drawn from
\begin{equation}
\label{eq:tensorbasis2}
\Gamma_0\in\left\{m_t \, \slashed{p}_3, \, m_t \, \slashed{p}_4\right\}\,, \qquad \Gamma_m\in\left\{m_t^2 \, \mathds{1}, \, m_t \, \slashed{p}_1, \, m_t \, \slashed{p}_2, \, \slashed{p}_1\slashed{p}_2\right\}\,,
    \qquad \kappa\in\left\{\varepsilon_5 \cdot p_1, \, \varepsilon_5 \cdot p_2, \, \varepsilon_5\cdot p_3\right\}\,.
\end{equation}
Here, $\overline{v}_1$ and $u_2$ denote the massless spinors associated with the momenta $p_1$ and $p_2$, respectively, while $\overline{V}_3$ and $U_4$ denote the corresponding massive spinors associated with $p_3$ and $p_4$.
The polarisation vector of the $W$ boson is denoted by $\varepsilon_5^\mu$.
Explicitly, the tensors are given by
\begingroup
\allowdisplaybreaks
\begin{alignat}{2}
    T_{1} &= m_t^3 \, \varepsilon_{5} \cdot p_1 \, \overline{v}_1 \slashed{p}_3 u_2 \, \overline{V}_3 U_4 \,,
    &\qquad \qquad
    T_{13} &= m_t^2 \, \varepsilon_{5} \cdot p_1 \, \overline{v}_1 \slashed{p}_3 u_2 \, \overline{V}_3 \slashed{p}_2 U_4\,,
    \nonumber\\
    T_{2} &= m_t^3 \, \varepsilon_{5} \cdot p_1 \, \overline{v}_1 \slashed{p}_4 u_2 \, \overline{V}_3 U_4\,,
    &\qquad \qquad
    T_{14} &= m_t^2 \, \varepsilon_{5} \cdot p_1 \, \overline{v}_1 \slashed{p}_4 u_2 \, \overline{V}_3 \slashed{p}_2 U_4\,,
    \nonumber\\
    T_{3} &= m_t^3 \, \varepsilon_{5} \cdot p_2 \, \overline{v}_1 \slashed{p}_3 u_2 \, \overline{V}_3 U_4\,,
    &\qquad \qquad
    T_{15} &= m_t^2 \, \varepsilon_{5} \cdot p_2 \, \overline{v}_1 \slashed{p}_3 u_2 \, \overline{V}_3 \slashed{p}_2 U_4\,,
    \nonumber\\
    T_{4} &= m_t^3 \, \varepsilon_{5} \cdot p_2 \, \overline{v}_1 \slashed{p}_4 u_2 \, \overline{V}_3 U_4\,,
    &\qquad \qquad
    T_{16} &= m_t^2 \, \varepsilon_{5} \cdot p_2 \, \overline{v}_1 \slashed{p}_4 u_2 \, \overline{V}_3 \slashed{p}_2 U_4\,,
    \nonumber\\
    T_{5} &= m_t^3 \, \varepsilon_{5} \cdot p_3 \, \overline{v}_1 \slashed{p}_3 u_2 \, \overline{V}_3 U_4\,,
    &\qquad \qquad
    T_{17} &= m_t^2 \, \varepsilon_{5} \cdot p_3 \, \overline{v}_1 \slashed{p}_3 u_2 \, \overline{V}_3 \slashed{p}_2 U_4\,,
    \nonumber\\
    T_{6} &= m_t^3 \, \varepsilon_{5} \cdot p_3 \, \overline{v}_1 \slashed{p}_4 u_2 \, \overline{V}_3 U_4\,,
    &\qquad \qquad
    T_{18} &= m_t^2 \, \varepsilon_{5} \cdot p_3 \, \overline{v}_1 \slashed{p}_4 u_2 \, \overline{V}_3 \slashed{p}_2 U_4\,,
    \\
    T_{7} &= m_t^2 \, \varepsilon_{5} \cdot p_1 \, \overline{v}_1 \slashed{p}_3 u_2 \, \overline{V}_3 \slashed{p}_1 U_4\,,
    &\qquad \qquad
    T_{19} &= m_t \, \varepsilon_{5} \cdot p_1 \, \overline{v}_1 \slashed{p}_3 u_2 \, \overline{V}_3 \slashed{p}_1 \slashed{p}_2 U_4\,,
    \nonumber\\
    T_{8} &= m_t^2 \, \varepsilon_{5} \cdot p_1 \, \overline{v}_1 \slashed{p}_4 u_2 \, \overline{V}_3 \slashed{p}_1 U_4\,,
    &\qquad \qquad
    T_{20} &= m_t \, \varepsilon_{5} \cdot p_1 \, \overline{v}_1 \slashed{p}_4 u_2 \, \overline{V}_3 \slashed{p}_1 \slashed{p}_2 U_4\,,
    \nonumber\\
    T_{9} &= m_t^2 \, \varepsilon_{5} \cdot p_2 \, \overline{v}_1 \slashed{p}_3 u_2 \, \overline{V}_3 \slashed{p}_1 U_4\,,
    &\qquad \qquad
    T_{21} &= m_t \, \varepsilon_{5} \cdot p_2 \, \overline{v}_1 \slashed{p}_3 u_2 \, \overline{V}_3 \slashed{p}_1 \slashed{p}_2 U_4\,,
    \nonumber\\
    T_{10} &= m_t^2 \, \varepsilon_{5} \cdot p_2 \, \overline{v}_1 \slashed{p}_4 u_2 \, \overline{V}_3 \slashed{p}_1 U_4\,,
    &\qquad \qquad
    T_{22} &= m_t \,  \varepsilon_{5} \cdot p_2 \,  \overline{v}_1 \slashed{p}_4 u_2 \, \overline{V}_3 \slashed{p}_1 \slashed{p}_2 U_4\,,
    \nonumber\\
    T_{11} &= m_t^2 \, \varepsilon_{5} \cdot p_3 \, \overline{v}_1 \slashed{p}_3 u_2 \, \overline{V}_3 \slashed{p}_1 U_4\,,
    &\qquad \qquad
    T_{23} &= m_t \, \varepsilon_{5} \cdot p_3 \, \overline{v}_1 \slashed{p}_3 u_2 \, \overline{V}_3 \slashed{p}_1 \slashed{p}_2 U_4\,,
    \nonumber\\
    T_{12} &= m_t^2 \, \varepsilon_{5} \cdot p_3 \, \overline{v}_1 \slashed{p}_4 u_2 \, \overline{V}_3 \slashed{p}_1 U_4\,,
    &\qquad \qquad
    T_{24} &= m_t \, \varepsilon_{5} \cdot p_3 \, \overline{v}_1 \slashed{p}_4 u_2 \, \overline{V}_3 \slashed{p}_1 \slashed{p}_2 U_4\, .
    \nonumber
\end{alignat}
\endgroup

\newpage

\bibliography{bibliography}

\providecommand{\href}[2]{#2}\begingroup\raggedright\begin{thebibliography}{10}

\bibitem{ATLAS:2018alq}
{\bf ATLAS} Collaboration, M.~Aaboud et~al., {\it {Search for new phenomena in
  events with same-charge leptons and $b$-jets in $pp$ collisions at $\sqrt{s}=
  13$ TeV with the ATLAS detector}},  {\em JHEP} {\bf 12} (2018) 039,
  [\href{http://arxiv.org/abs/1807.11883}{{\tt arXiv:1807.11883}}].

\bibitem{ATLAS:2019fag}
{\bf ATLAS} Collaboration, G.~Aad et~al., {\it {Search for squarks and gluinos
  in final states with same-sign leptons and jets using 139 fb$^{-1}$ of data
  collected with the ATLAS detector}},  {\em JHEP} {\bf 06} (2020) 046,
  [\href{http://arxiv.org/abs/1909.08457}{{\tt arXiv:1909.08457}}].

\bibitem{CMS:2020cpy}
{\bf CMS} Collaboration, A.~M. Sirunyan et~al., {\it {Search for physics beyond
  the standard model in events with jets and two same-sign or at least three
  charged leptons in proton-proton collisions at $\sqrt{s}=$ 13 TeV}},  {\em
  Eur. Phys. J. C} {\bf 80} (2020), no.~8 752,
  [\href{http://arxiv.org/abs/2001.10086}{{\tt arXiv:2001.10086}}].

\bibitem{CMS:2022tkv}
{\bf CMS} Collaboration, A.~Tumasyan et~al., {\it {Measurement of the cross
  section of top quark-antiquark pair production in association with a W boson
  in proton-proton collisions at $\sqrt{s} $ = 13 TeV}},  {\em JHEP} {\bf 07}
  (2023) 219, [\href{http://arxiv.org/abs/2208.06485}{{\tt arXiv:2208.06485}}].

\bibitem{ATLAS:2024moy}
{\bf ATLAS} Collaboration, G.~Aad et~al., {\it {Measurement of the total and
  differential cross-sections of $ t\overline{t}W $ production in pp collisions
  at $ \sqrt{s} $ = 13 TeV with the ATLAS detector}},  {\em JHEP} {\bf 05}
  (2024) 131, [\href{http://arxiv.org/abs/2401.05299}{{\tt arXiv:2401.05299}}].
  [Erratum: JHEP 11, 127 (2025)].

\bibitem{CMS:2025iwa}
{\bf CMS} Collaboration, A.~Hayrapetyan et~al., {\it {Measurements of $
  \textrm{t}\overline{\textrm{t}}\textrm{W} $ differential cross sections and
  the leptonic charge asymmetry at $ \sqrt{s}=13 $ TeV}},  {\em JHEP} {\bf 03}
  (2026) 083, [\href{http://arxiv.org/abs/2509.13512}{{\tt arXiv:2509.13512}}].

\bibitem{ATLAS:2019nvo}
{\bf ATLAS} Collaboration, {\it {Analysis of $t\bar{t}H$ and $t\bar{t}W$
  production in multilepton final states with the ATLAS detector}},  tech.
  rep., CERN, Geneva, 2019.

\bibitem{CMS:2020mpn}
{\bf CMS} Collaboration, A.~M. Sirunyan et~al., {\it {Measurement of the Higgs
  boson production rate in association with top quarks in final states with
  electrons, muons, and hadronically decaying tau leptons at $\sqrt{s} =$ 13
  TeV}},  {\em Eur. Phys. J. C} {\bf 81} (2021), no.~4 378,
  [\href{http://arxiv.org/abs/2011.03652}{{\tt arXiv:2011.03652}}].

\bibitem{CMS:2023ftu}
{\bf CMS} Collaboration, A.~Hayrapetyan et~al., {\it {Observation of four top
  quark production in proton-proton collisions at $ \sqrt{s}=13 $ TeV}},  {\em
  Phys. Lett. B} {\bf 847} (2023) 138290,
  [\href{http://arxiv.org/abs/2305.13439}{{\tt arXiv:2305.13439}}].

\bibitem{ATLAS:2023ajo}
{\bf ATLAS} Collaboration, G.~Aad et~al., {\it {Observation of four-top-quark
  production in the multilepton final state with the ATLAS detector}},  {\em
  Eur. Phys. J. C} {\bf 83} (2023), no.~6 496,
  [\href{http://arxiv.org/abs/2303.15061}{{\tt arXiv:2303.15061}}]. [Erratum:
  Eur.Phys.J.C 84, 156 (2024)].

\bibitem{Badger:2010mg}
S.~Badger, J.~M. Campbell, and R.~K. Ellis, {\it {QCD Corrections to the
  Hadronic Production of a Heavy Quark Pair and a W-Boson Including Decay
  Correlations}},  {\em JHEP} {\bf 03} (2011) 027,
  [\href{http://arxiv.org/abs/1011.6647}{{\tt arXiv:1011.6647}}].

\bibitem{Campbell:2012dh}
J.~M. Campbell and R.~K. Ellis, {\it {$t \bar{t} W^{+-}$ production and decay
  at NLO}},  {\em JHEP} {\bf 07} (2012) 052,
  [\href{http://arxiv.org/abs/1204.5678}{{\tt arXiv:1204.5678}}].

\bibitem{Maltoni:2015ena}
F.~Maltoni, D.~Pagani, and I.~Tsinikos, {\it {Associated production of a
  top-quark pair with vector bosons at NLO in QCD: impact on $
  \mathrm{t}\overline{\mathrm{t}}\mathrm{H} $ searches at the LHC}},  {\em
  JHEP} {\bf 02} (2016) 113, [\href{http://arxiv.org/abs/1507.05640}{{\tt
  arXiv:1507.05640}}].

\bibitem{Frixione:2015zaa}
S.~Frixione, V.~Hirschi, D.~Pagani, H.~S. Shao, and M.~Zaro, {\it {Electroweak
  and QCD corrections to top-pair hadroproduction in association with heavy
  bosons}},  {\em JHEP} {\bf 06} (2015) 184,
  [\href{http://arxiv.org/abs/1504.03446}{{\tt arXiv:1504.03446}}].

\bibitem{Frederix:2017wme}
R.~Frederix, D.~Pagani, and M.~Zaro, {\it {Large NLO corrections in
  $t\bar{t}W^{\pm}$ and $t\bar{t}t\bar{t}$ hadroproduction from supposedly
  subleading EW contributions}},  {\em JHEP} {\bf 02} (2018) 031,
  [\href{http://arxiv.org/abs/1711.02116}{{\tt arXiv:1711.02116}}].

\bibitem{Bevilacqua:2020pzy}
G.~Bevilacqua, H.-Y. Bi, H.~B. Hartanto, M.~Kraus, and M.~Worek, {\it {The
  simplest of them all: $t\bar{t} W^\pm$ at NLO accuracy in QCD}},  {\em JHEP}
  {\bf 08} (2020) 043, [\href{http://arxiv.org/abs/2005.09427}{{\tt
  arXiv:2005.09427}}].

\bibitem{Denner:2020hgg}
A.~Denner and G.~Pelliccioli, {\it {NLO QCD corrections to off-shell
  $\text{t}\bar{\text{t}}\text{W}^+$ production at the LHC}},  {\em JHEP} {\bf
  11} (2020) 069, [\href{http://arxiv.org/abs/2007.12089}{{\tt
  arXiv:2007.12089}}].

\bibitem{Bevilacqua:2020srb}
G.~Bevilacqua, H.-Y. Bi, H.~B. Hartanto, M.~Kraus, J.~Nasufi, and M.~Worek,
  {\it {NLO QCD corrections to off-shell ${t{\bar{t}}W^\pm }$ production at the
  LHC: correlations and asymmetries}},  {\em Eur. Phys. J. C} {\bf 81} (2021),
  no.~7 675, [\href{http://arxiv.org/abs/2012.01363}{{\tt arXiv:2012.01363}}].

\bibitem{Denner:2021hqi}
A.~Denner and G.~Pelliccioli, {\it {Combined NLO EW and QCD corrections to
  off-shell $\text {t} \overline{\text {t}}\text {W} $ production at the LHC}},
   {\em Eur. Phys. J. C} {\bf 81} (2021), no.~4 354,
  [\href{http://arxiv.org/abs/2102.03246}{{\tt arXiv:2102.03246}}].

\bibitem{Frederix:2012ps}
R.~Frederix and S.~Frixione, {\it {Merging meets matching in MC@NLO}},  {\em
  JHEP} {\bf 12} (2012) 061, [\href{http://arxiv.org/abs/1209.6215}{{\tt
  arXiv:1209.6215}}].

\bibitem{Frederix:2021agh}
R.~Frederix and I.~Tsinikos, {\it {On improving NLO merging for $
  \mathrm{t}\overline{\mathrm{t}}\mathrm{W} $ production}},  {\em JHEP} {\bf
  11} (2021) 029, [\href{http://arxiv.org/abs/2108.07826}{{\tt
  arXiv:2108.07826}}].

\bibitem{Dimitrakopoulos:2026jwi}
N.~Dimitrakopoulos and M.~Worek, {\it {Multi-scale improved predictions for $pp
  \to t\bar{t}W^+ +X$}},  \href{http://arxiv.org/abs/2607.11652}{{\tt
  arXiv:2607.11652}}.

\bibitem{Buonocore:2023ljm}
L.~Buonocore, S.~Devoto, M.~Grazzini, S.~Kallweit, J.~Mazzitelli, L.~Rottoli,
  and C.~Savoini, {\it {Precise Predictions for the Associated Production of a
  W Boson with a Top-Antitop Quark Pair at the LHC}},  {\em Phys. Rev. Lett.}
  {\bf 131} (2023), no.~23 231901, [\href{http://arxiv.org/abs/2306.16311}{{\tt
  arXiv:2306.16311}}].

\bibitem{FebresCordero:2023pww}
F.~Febres~Cordero, G.~Figueiredo, M.~Kraus, B.~Page, and L.~Reina, {\it
  {Two-loop master integrals for leading-color $ pp\to t\overline{t}H $
  amplitudes with a light-quark loop}},  {\em JHEP} {\bf 07} (2024) 084,
  [\href{http://arxiv.org/abs/2312.08131}{{\tt arXiv:2312.08131}}].

\bibitem{Agarwal:2024jyq}
B.~Agarwal, G.~Heinrich, S.~P. Jones, M.~Kerner, S.~Y. Klein, J.~Lang,
  V.~Magerya, and A.~Olsson, {\it {Two-loop amplitudes for $ t\overline{t}H $
  production: the quark-initiated N$_{f}$-part}},  {\em JHEP} {\bf 05} (2024)
  013, [\href{http://arxiv.org/abs/2402.03301}{{\tt arXiv:2402.03301}}].
  [Erratum: JHEP 06, 142 (2024)].

\bibitem{Badger:2025ljy}
S.~Badger, M.~Becchetti, C.~Brancaccio, M.~Czakon, H.~B. Hartanto, R.~Poncelet,
  and S.~Zoia, {\it {Double virtual QCD corrections to $ t\overline{t} $+jet
  production at the LHC}},  {\em JHEP} {\bf 05} (2026) 044,
  [\href{http://arxiv.org/abs/2511.11424}{{\tt arXiv:2511.11424}}].

\bibitem{Badger:2024dxo}
S.~Badger, M.~Becchetti, C.~Brancaccio, H.~B. Hartanto, and S.~Zoia, {\it
  {Numerical evaluation of two-loop QCD helicity amplitudes for $ gg\to
  t\overline{t}g $ at leading colour}},  {\em JHEP} {\bf 03} (2025) 070,
  [\href{http://arxiv.org/abs/2412.13876}{{\tt arXiv:2412.13876}}].

\bibitem{Becchetti:2025qlu}
M.~Becchetti, D.~Canko, V.~Chestnov, T.~Peraro, M.~Pozzoli, and S.~Zoia, {\it
  {Two-loop Feynman integrals for leading colour $ t\overline{t}W $ production
  at hadron colliders}},  {\em JHEP} {\bf 07} (2025) 001,
  [\href{http://arxiv.org/abs/2504.13011}{{\tt arXiv:2504.13011}}].

\bibitem{Gehrmann:2018yef}
T.~Gehrmann, J.~M. Henn, and N.~A. Lo~Presti, {\it {Pentagon functions for
  massless planar scattering amplitudes}},  {\em JHEP} {\bf 10} (2018) 103,
  [\href{http://arxiv.org/abs/1807.09812}{{\tt arXiv:1807.09812}}].

\bibitem{Chicherin:2020oor}
D.~Chicherin and V.~Sotnikov, {\it {Pentagon Functions for Scattering of Five
  Massless Particles}},  {\em JHEP} {\bf 20} (2020) 167,
  [\href{http://arxiv.org/abs/2009.07803}{{\tt arXiv:2009.07803}}].

\bibitem{Chicherin:2021dyp}
D.~Chicherin, V.~Sotnikov, and S.~Zoia, {\it {Pentagon functions for one-mass
  planar scattering amplitudes}},  {\em JHEP} {\bf 01} (2022) 096,
  [\href{http://arxiv.org/abs/2110.10111}{{\tt arXiv:2110.10111}}].

\bibitem{Abreu:2023rco}
S.~Abreu, D.~Chicherin, H.~Ita, B.~Page, V.~Sotnikov, W.~Tschernow, and
  S.~Zoia, {\it {All Two-Loop Feynman Integrals for Five-Point One-Mass
  Scattering}},  {\em Phys. Rev. Lett.} {\bf 132} (2024), no.~14 141601,
  [\href{http://arxiv.org/abs/2306.15431}{{\tt arXiv:2306.15431}}].

\bibitem{Henn:2013pwa}
J.~M. Henn, {\it {Multiloop integrals in dimensional regularization made
  simple}},  {\em Phys. Rev. Lett.} {\bf 110} (2013) 251601,
  [\href{http://arxiv.org/abs/1304.1806}{{\tt arXiv:1304.1806}}].

\bibitem{Liu:2022chg}
X.~Liu and Y.-Q. Ma, {\it {AMFlow: A Mathematica package for Feynman integrals
  computation via auxiliary mass flow}},  {\em Comput. Phys. Commun.} {\bf 283}
  (2023) 108565, [\href{http://arxiv.org/abs/2201.11669}{{\tt
  arXiv:2201.11669}}].

\bibitem{vonManteuffel:2014ixa}
A.~von Manteuffel and R.~M. Schabinger, {\it {A novel approach to integration
  by parts reduction}},  {\em Phys. Lett. B} {\bf 744} (2015) 101--104,
  [\href{http://arxiv.org/abs/1406.4513}{{\tt arXiv:1406.4513}}].

\bibitem{Peraro:2016wsq}
T.~Peraro, {\it {Scattering amplitudes over finite fields and multivariate
  functional reconstruction}},  {\em JHEP} {\bf 12} (2016) 030,
  [\href{http://arxiv.org/abs/1608.01902}{{\tt arXiv:1608.01902}}].

\bibitem{Becchetti:2026awn}
M.~Becchetti et~al., {\it {NNLO QCD predictions for $t\bar t W$ production at
  hadron colliders}},  \href{http://arxiv.org/abs/2606.09503}{{\tt
  arXiv:2606.09503}}.

\bibitem{Peraro:2019okx}
T.~Peraro, {\it {Analytic multi-loop results using finite fields and dataflow
  graphs with FiniteFlow}},  in {\em {14th International Symposium on Radiative
  Corrections}: {Application of Quantum Field Theory to Phenomenology}}, 12,
  2019.
\newblock \href{http://arxiv.org/abs/1912.03142}{{\tt arXiv:1912.03142}}.

\bibitem{Becchetti:2025osw}
M.~Becchetti, M.~Delto, S.~Ditsch, P.~A. Kreer, M.~Pozzoli, and L.~Tancredi,
  {\it {One-loop QCD corrections to $\bar{u} d \to t\bar{t}W $ at
  $\mathcal{O}\left({\varepsilon}^2\right) $}},  {\em JHEP} {\bf 09} (2025)
  126, [\href{http://arxiv.org/abs/2502.14952}{{\tt arXiv:2502.14952}}].

\bibitem{Catani:1998bh}
S.~Catani, {\it {The Singular behavior of QCD amplitudes at two loop order}},
  {\em Phys. Lett. B} {\bf 427} (1998) 161--171,
  [\href{http://arxiv.org/abs/hep-ph/9802439}{{\tt hep-ph/9802439}}].

\bibitem{Gardi:2009qi}
E.~Gardi and L.~Magnea, {\it {Factorization constraints for soft anomalous
  dimensions in QCD scattering amplitudes}},  {\em JHEP} {\bf 03} (2009) 079,
  [\href{http://arxiv.org/abs/0901.1091}{{\tt arXiv:0901.1091}}].

\bibitem{Gardi:2009zv}
E.~Gardi and L.~Magnea, {\it {Infrared singularities in QCD amplitudes}},  {\em
  Nuovo Cim. C} {\bf 32N5-6} (2009) 137--157,
  [\href{http://arxiv.org/abs/0908.3273}{{\tt arXiv:0908.3273}}].

\bibitem{Becher:2009cu}
T.~Becher and M.~Neubert, {\it {Infrared singularities of scattering amplitudes
  in perturbative QCD}},  {\em Phys. Rev. Lett.} {\bf 102} (2009) 162001,
  [\href{http://arxiv.org/abs/0901.0722}{{\tt arXiv:0901.0722}}]. [Erratum:
  Phys.Rev.Lett. 111, 199905 (2013)].

\bibitem{Becher:2009qa}
T.~Becher and M.~Neubert, {\it {On the Structure of Infrared Singularities of
  Gauge-Theory Amplitudes}},  {\em JHEP} {\bf 06} (2009) 081,
  [\href{http://arxiv.org/abs/0903.1126}{{\tt arXiv:0903.1126}}]. [Erratum:
  JHEP 11, 024 (2013)].

\bibitem{Becher:2009kw}
T.~Becher and M.~Neubert, {\it {Infrared singularities of QCD amplitudes with
  massive partons}},  {\em Phys. Rev. D} {\bf 79} (2009) 125004,
  [\href{http://arxiv.org/abs/0904.1021}{{\tt arXiv:0904.1021}}]. [Erratum:
  Phys.Rev.D 80, 109901 (2009)].

\bibitem{Ferroglia:2009ep}
A.~Ferroglia, M.~Neubert, B.~D. Pecjak, and L.~L. Yang, {\it {Two-loop
  divergences of scattering amplitudes with massive partons}},  {\em Phys. Rev.
  Lett.} {\bf 103} (2009) 201601, [\href{http://arxiv.org/abs/0907.4791}{{\tt
  arXiv:0907.4791}}].

\bibitem{Ferroglia:2009ii}
A.~Ferroglia, M.~Neubert, B.~D. Pecjak, and L.~L. Yang, {\it {Two-loop
  divergences of massive scattering amplitudes in non-abelian gauge theories}},
   {\em JHEP} {\bf 11} (2009) 062, [\href{http://arxiv.org/abs/0908.3676}{{\tt
  arXiv:0908.3676}}].

\bibitem{tHooft:1972tcz}
G.~'t~Hooft and M.~J.~G. Veltman, {\it {Regularization and Renormalization of
  Gauge Fields}},  {\em Nucl. Phys. B} {\bf 44} (1972) 189--213.

\bibitem{Peraro:2019cjj}
T.~Peraro and L.~Tancredi, {\it {Physical projectors for multi-leg helicity
  amplitudes}},  {\em JHEP} {\bf 07} (2019) 114,
  [\href{http://arxiv.org/abs/1906.03298}{{\tt arXiv:1906.03298}}].

\bibitem{Peraro:2020sfm}
T.~Peraro and L.~Tancredi, {\it {Tensor decomposition for bosonic and fermionic
  scattering amplitudes}},  {\em Phys. Rev. D} {\bf 103} (2021), no.~5 054042,
  [\href{http://arxiv.org/abs/2012.00820}{{\tt arXiv:2012.00820}}].

\bibitem{Nogueira:1991ex}
P.~Nogueira, {\it {Automatic Feynman Graph Generation}},  {\em J. Comput.
  Phys.} {\bf 105} (1993) 279--289.

\bibitem{Ruijl:2017dtg}
B.~Ruijl, T.~Ueda, and J.~Vermaseren, {\it {FORM version 4.2}},
  \href{http://arxiv.org/abs/1707.06453}{{\tt arXiv:1707.06453}}.

\bibitem{vonManteuffel:2012np}
A.~von Manteuffel and C.~Studerus, {\it {Reduze 2 - Distributed Feynman
  Integral Reduction}},  \href{http://arxiv.org/abs/1201.4330}{{\tt
  arXiv:1201.4330}}.

\bibitem{Tkachov:1981wb}
F.~V. Tkachov, {\it {A theorem on analytical calculability of 4-loop
  renormalization group functions}},  {\em Phys. Lett. B} {\bf 100} (1981)
  65--68.

\bibitem{Chetyrkin:1981qh}
K.~G. Chetyrkin and F.~V. Tkachov, {\it {Integration by parts: The algorithm to
  calculate $\beta$-functions in 4 loops}},  {\em Nucl. Phys. B} {\bf 192}
  (1981) 159--204.

\bibitem{Laporta:2000dsw}
S.~Laporta, {\it {High-precision calculation of multiloop Feynman integrals by
  difference equations}},  {\em Int. J. Mod. Phys. A} {\bf 15} (2000)
  5087--5159, [\href{http://arxiv.org/abs/hep-ph/0102033}{{\tt
  hep-ph/0102033}}].

\bibitem{Wu:2023upw}
Z.~Wu, J.~Boehm, R.~Ma, H.~Xu, and Y.~Zhang, {\it {NeatIBP 1.0, a package
  generating small-size integration-by-parts relations for Feynman integrals}},
   {\em Comput. Phys. Commun.} {\bf 295} (2024) 108999,
  [\href{http://arxiv.org/abs/2305.08783}{{\tt arXiv:2305.08783}}].

\bibitem{Peraro:2019svx}
T.~Peraro, {\it {$\text{FiniteFlow}$: multivariate functional reconstruction
  using finite fields and dataflow graphs}},  {\em JHEP} {\bf 07} (2019) 031,
  [\href{http://arxiv.org/abs/1905.08019}{{\tt arXiv:1905.08019}}].

\bibitem{Badger:2021imn}
S.~Badger, C.~Br{\o}nnum-Hansen, D.~Chicherin, T.~Gehrmann, H.~B. Hartanto,
  J.~Henn, M.~Marcoli, R.~Moodie, T.~Peraro, and S.~Zoia, {\it {Virtual QCD
  corrections to gluon-initiated diphoton plus jet production at hadron
  colliders}},  {\em JHEP} {\bf 11} (2021) 083,
  [\href{http://arxiv.org/abs/2106.08664}{{\tt arXiv:2106.08664}}].

\bibitem{Wang:1981:PAU:800206.806398}
P.~S. Wang, {\it A p-adic algorithm for univariate partial fractions},  in {\em
  Proceedings of the Fourth ACM Symposium on Symbolic and Algebraic
  Computation}, SYMSAC '81, (New York, NY, USA), pp.~212--217, ACM, 1981.

\bibitem{Wang:1982:PRR:1089292.1089293}
P.~S. Wang, M.~J.~T. Guy, and J.~H. Davenport, {\it P-adic reconstruction of
  rational numbers},  {\em SIGSAM Bull.} {\bf 16} (May, 1982) 2--3.

\bibitem{Abreu:2018zmy}
S.~Abreu, J.~Dormans, F.~Febres~Cordero, H.~Ita, and B.~Page, {\it {Analytic
  Form of Planar Two-Loop Five-Gluon Scattering Amplitudes in QCD}},  {\em
  Phys. Rev. Lett.} {\bf 122} (2019), no.~8 082002,
  [\href{http://arxiv.org/abs/1812.04586}{{\tt arXiv:1812.04586}}].

\bibitem{Liu:2017jxz}
X.~Liu, Y.-Q. Ma, and C.-Y. Wang, {\it {A Systematic and Efficient Method to
  Compute Multi-loop Master Integrals}},  {\em Phys. Lett. B} {\bf 779} (2018)
  353--357, [\href{http://arxiv.org/abs/1711.09572}{{\tt arXiv:1711.09572}}].

\bibitem{Liu:2021wks}
X.~Liu and Y.-Q. Ma, {\it {Multiloop corrections for collider processes using
  auxiliary mass flow}},  {\em Phys. Rev. D} {\bf 105} (2022), no.~5 L051503,
  [\href{http://arxiv.org/abs/2107.01864}{{\tt arXiv:2107.01864}}].

\bibitem{Chen:1977oja}
K.-T. Chen, {\it {Iterated path integrals}},  {\em Bull. Am. Math. Soc.} {\bf
  83} (1977) 831--879.

\bibitem{Goncharov:2010jf}
A.~B. Goncharov, M.~Spradlin, C.~Vergu, and A.~Volovich, {\it {Classical
  Polylogarithms for Amplitudes and Wilson Loops}},  {\em Phys. Rev. Lett.}
  {\bf 105} (2010) 151605, [\href{http://arxiv.org/abs/1006.5703}{{\tt
  arXiv:1006.5703}}].

\bibitem{pslq}
H.~Ferguson and D.~Bailey, {\it {A Polynomial Time, Numerically Stable Integer
  Relation Algorithm}},  {\em RNR Technical Report RNR-91-032} (1992).

\bibitem{zenodo}
M.~Becchetti, D.~Canko, X.~Chen, V.~Chestnov, M.~Delto, S.~Ditsch, T.~Peraro,
  M.~Pozzoli, C.~Savoini, and S.~Zoia, {\it {Ancillary files for ``Two-loop QCD
  amplitudes for $t\bar{t}W$ production at the LHC in the leading-colour
  approximation''}},  8, 2026.
\newblock {DOI}
  \href{https://doi.org/10.5281/zenodo.21741495}{10.5281/zenodo.21741495}.

\bibitem{Lee:2013mka}
R.~N. Lee, {\it {LiteRed 1.4: a powerful tool for reduction of multiloop
  integrals}},  {\em J. Phys. Conf. Ser.} {\bf 523} (2014) 012059,
  [\href{http://arxiv.org/abs/1310.1145}{{\tt arXiv:1310.1145}}].

\bibitem{Pozzorini:2005ff}
S.~Pozzorini and E.~Remiddi, {\it {Precise numerical evaluation of the two loop
  sunrise graph master integrals in the equal mass case}},  {\em Comput. Phys.
  Commun.} {\bf 175} (2006) 381--387,
  [\href{http://arxiv.org/abs/hep-ph/0505041}{{\tt hep-ph/0505041}}].

\bibitem{Moriello:2019yhu}
F.~Moriello, {\it {Generalised power series expansions for the elliptic planar
  families of Higgs + jet production at two loops}},  {\em JHEP} {\bf 01}
  (2020) 150, [\href{http://arxiv.org/abs/1907.13234}{{\tt arXiv:1907.13234}}].

\bibitem{Costantini:2024wby}
M.~N. Costantini, M.~Madigan, L.~Mantani, and J.~M. Moore, {\it {A critical
  study of the Monte Carlo replica method}},  {\em JHEP} {\bf 12} (2024) 064,
  [\href{http://arxiv.org/abs/2404.10056}{{\tt arXiv:2404.10056}}].

\bibitem{Buccioni:2019sur}
F.~Buccioni, J.-N. Lang, J.~M. Lindert, P.~Maierh{\"o}fer, S.~Pozzorini,
  H.~Zhang, and M.~F. Zoller, {\it {OpenLoops 2}},  {\em Eur. Phys. J. C} {\bf
  79} (2019), no.~10 866, [\href{http://arxiv.org/abs/1907.13071}{{\tt
  arXiv:1907.13071}}].

\bibitem{Hahn:2000kx}
T.~Hahn, {\it {Generating Feynman diagrams and amplitudes with FeynArts 3}},
  {\em Comput. Phys. Commun.} {\bf 140} (2001) 418--431,
  [\href{http://arxiv.org/abs/hep-ph/0012260}{{\tt hep-ph/0012260}}].

\bibitem{Bi:2023bnq}
H.-Y. Bi, L.-H. Huang, R.-J. Huang, Y.-Q. Ma, and H.-M. Yu, {\it {Electroweak
  Corrections to Double Higgs Production at the LHC}},  {\em Phys. Rev. Lett.}
  {\bf 132} (2024), no.~23 231802, [\href{http://arxiv.org/abs/2311.16963}{{\tt
  arXiv:2311.16963}}].

\bibitem{Bi:2025oga}
H.-Y. Bi, Y.-Q. Ma, and D.-M. Mu, {\it {Electroweak loop corrections to gg
  {\textrightarrow} gH at the LHC}},  {\em JHEP} {\bf 04} (2026) 087,
  [\href{http://arxiv.org/abs/2508.02588}{{\tt arXiv:2508.02588}}].

\bibitem{Li:2025bsq}
H.~T. Li, Y.-Q. Ma, C.-T. Tan, J.~Wang, and H.-F. Zhang, {\it
  {Compton-Scattering Total Cross Section at Next-to-Next-to-Leading Order and
  Resummation of Leading Logarithms}},  {\em Phys. Rev. Lett.} {\bf 136}
  (2026), no.~2 021802, [\href{http://arxiv.org/abs/2511.09330}{{\tt
  arXiv:2511.09330}}].

\bibitem{Guan:2024byi}
X.~Guan, X.~Liu, Y.-Q. Ma, and W.-H. Wu, {\it {Blade: A package for
  block-triangular form improved Feynman integrals decomposition}},  {\em
  Comput. Phys. Commun.} {\bf 310} (2025) 109538,
  [\href{http://arxiv.org/abs/2405.14621}{{\tt arXiv:2405.14621}}].

\bibitem{Guan:2019bcx}
X.~Guan, X.~Liu, and Y.-Q. Ma, {\it {Complete reduction of integrals in
  two-loop five-light-parton scattering amplitudes}},  {\em Chin. Phys. C} {\bf
  44} (2020), no.~9 093106, [\href{http://arxiv.org/abs/1912.09294}{{\tt
  arXiv:1912.09294}}].

\bibitem{Breso-Pla:2024pda}
V.~Bres{\'o}-Pla, G.~Heinrich, V.~Magerya, and A.~Olsson, {\it {Interpolating
  amplitudes}},  {\em SciPost Phys.} {\bf 19} (2025), no.~5 123,
  [\href{http://arxiv.org/abs/2412.09534}{{\tt arXiv:2412.09534}}].

\bibitem{Broadhurst:1991fy}
D.~J. Broadhurst, N.~Gray, and K.~Schilcher, {\it {Gauge invariant on-shell
  Z(2) in QED, QCD and the effective field theory of a static quark}},  {\em Z.
  Phys. C} {\bf 52} (1991) 111--122.

\bibitem{Barnreuther:2013qvf}
P.~B{\"a}rnreuther, M.~Czakon, and P.~Fiedler, {\it {Virtual amplitudes and
  threshold behaviour of hadronic top-quark pair-production cross sections}},
  {\em JHEP} {\bf 02} (2014) 078, [\href{http://arxiv.org/abs/1312.6279}{{\tt
  arXiv:1312.6279}}].

\end{thebibliography}\endgroup

\end{document}